\documentclass{aa}
\usepackage{txfonts}
\usepackage{natbib,twoopt}
\usepackage{longtable}
\usepackage{graphicx}
\usepackage{threeparttable}
\usepackage{setspace}
\usepackage{arydshln}
\usepackage{hyperref}
\bibpunct{(}{)}{;}{a}{}{,} 

\hypersetup{
	pdftitle={A three-dimensional hydrodynamical line profile analysis of iron lines and barium isotopes in HD\,140283},
	pdfauthor={A. J. Gallagher},
	pdfsubject={A\&A journal},
	pdfcreator={A. J. Gallagher},
		colorlinks, 
		citecolor=blue, 
		filecolor=blue, 
		linkcolor=blue,
		breaklinks=true, 
		plainpages=false,
		urlcolor=blue 
}

\title{A three-dimensional hydrodynamical line profile analysis of iron lines and barium isotopes in HD\,140283\thanks{Based on observations taken using the Subaru Telescope High Dispersion Spectrograph (HDS), operated by the National Astronomical Observatory of Japan.}${}^{,}$\thanks{Colour figures are only available through access of the online journal.}}

	\author{A. J. Gallagher\inst{1,2}
\and
H.-G. Ludwig\inst{3,2}
\and
S. G. Ryan\inst{1}
\and
W. Aoki\inst{4}
}

\institute{Centre for Astrophysics Research, School of Physics, Astronomy \& Mathematics, University of Hertfordshire, College Lane, Hatfield, Hertfordshire, AL10 9AB, United Kingdom.\\ 
email: \texttt{andrew.gallagher@obspm.fr} 
\and GEPI, Observatoire de Paris, CNRS, Universit{\'e} Paris Diderot, Place Jules Janssen, 92190 Meudon, France.
\and Zentrum f{\"u}r Astrononmie der Universit{\"a}t Heidelberg, Landessternwarte, K{\"o}nigstuhl 12, 69117 Heidelberg, Germany.
\and National Astronomical Observatory, Mitaka, Tokyo, 181-8588, Japan.
}

\date{Received 14 August 2014 / Accepted 1 April 2015 }

\authorrunning{A. J. Gallagher et al.}
\titlerunning{A 3D line profile analysis of iron lines and barium isotopes in HD\,140283}

\newcommand{\Teff}{T_{\rm eff}}
\newcommand{\kms}{\rm km\,s^{-1}}
\newcommand{\fodd}{f_{\rm odd}}

\newcommand{\logg}{\log{g}}
\newcommand{\loggf}{\log{gf}}
\newcommand{\Msol}{{\rm M}_{\sun}}
\newcommand{\rsun}{{\rm R}_{\sun}}

\newcommand{\afe}{A(\element{Fe})}
\newcommand{\aba}{A(\element{Ba})}

\newcommand{\po}{{\footnotesize PAPER1}}
\newcommand{\atlas}{{\footnotesize ATLAS}}
\newcommand{\kurucz}{{\footnotesize KURUCZ06}}
\newcommand{\linfor}{{Linfor3D}}
\newcommand{\cobold}{{\footnotesize CO${}^{5}$BOLD}}
\newcommand{\od}{{\tt LHD}}
\newcommand{\vsini}{{\upsilon}\sin{i}}
\newcommand{\tda}{$\langle{\rm 3D}\rangle$}
\newcommand{\fei}{\ion{Fe}{I}}
\newcommand{\feii}{\ion{Fe}{II}}
\newcommand{\baii}{\ion{Ba}{II}}
\newcommand{\ltaur}{\log{\tau_{\rm ROSS}}}

\abstract{Heavy-elements, i.e. those beyond the iron peak, mostly form via two neutron capture processes: the slow (s-) and the rapid (r-) process. Metal-poor stars should contain fewer isotopes that form via the s-process, according to currently accepted theory. It has been shown in several investigations that theory and observation do not agree well, raising questions on the validity of either the methodology or the theory.}
{We analyse the metal-poor star HD\,140283, for which we have a high quality spectrum. We test whether a three-dimensional (3D) local thermodynamic equilibrium (LTE) stellar atmosphere and spectrum synthesis code permits a more reliable analysis of the iron abundance and barium isotope ratio than a one-dimensional (1D) LTE analysis.}
{Using 3D hydrodynamical model atmospheres, we examine 91 iron lines of varying strength and formation depth. This provides us with the star's rotational speed. With this, we model the barium isotope ratio by exploiting the hyperfine structure of the singly ionised 4554\,\AA\ resonance line, and study the impact of the uncertainties in the stellar parameters.}
{The star's rotational speed was found to be $1.65\pm0.05\,\kms$. Barium isotopes under the 3D paradigm show a dominant r-process signature as $77\pm6\pm17\%$ ($\fodd=0.38\pm0.02\pm0.06$) of barium isotopes form via the r-process, where errors represent the assigned random and systematic errors, respectively. We find that 3D LTE fits reproduce iron line profiles better than those in 1D, but do not provide a unique abundance (within the uncertainties). However, we demonstrate that the isotopic ratio is robust against this shortcoming.}
{Our barium isotope result agrees well with currently accepted theory regarding the formation of the heavy-elements during the early Galaxy. The improved fit to the asymmetric iron line profiles suggests that the current state of 3D LTE modelling provides excellent simulations of fluid flows. However, the abundances they provide are not yet self-consistent. This may improve with non-local thermodynamic equilibrium considerations and higher resolution models.}

\begin{document}

\maketitle
\section{Introduction}
\label{sec:introduction}

Traditional equivalent width analyses of stellar spectra have employed one-dimensional (1D) model atmospheres and ignored the details of the line profile shapes. However, this ignores valuable information on the state of the stellar atmosphere, and for some investigations it is the line profile that harbours the scientific reward. In this paper we test whether one class of three-dimensional (3D) models are currently better able to model iron line profiles in a metal-poor subgiant and consider the barium isotope ratio they imply.

The nearby subgiant star HD\,140283 is one of the brightest metal-poor stars \citep[$V=7.21$,][]{Casagrande2010} and as a consequence of the high signal-to-noise ($S/N$) and resolving power that is achievable, it has been the subject of extensive spectroscopic study. Its age and origin \citep{Bond2013} and its composition have been independently determined in a plethora of publications \citep[e.g.][]{Magain1993,Bonifacio1998,Thoren2000,Asplund2003,Shchukina2005,Roederer2012,Lind2012}. It is one of the most well-studied metal-poor stars. 

There have been numerous analyses of HD\,140283's neutron-capture element abundances \citep[e.g.][]{Peterson2011,Roederer2012,Mello2012}. By measuring the ratio of a heavy element whose origin is primarily of the rapid (r-) neutron-capture process with another synthesised primarily through the slow (s-) process, it is possible to determine the ratio of r- to s-process in a star. A typical example of this technique is the measurement of europium and barium ratios, where 94\% and 81\% of the solar system abundance can be traced through the r- and s-process, respectively \citep{Arlandini1999}. Using theoretical nuclear networks such as \citet{Arlandini1999} or \citet{Burris2000} allows one to put limits on the r- and s-process ratio. Using the \citet{Arlandini1999} nuclear network, the s-process-only [Ba/Eu]\footnote{${\rm [X/Y]}=\log_{10}{\left({\frac{N({\rm X})}{N({\rm Y})}}\right)_{*}} - \log_{10}{\left({\frac{N({\rm X})}{N({\rm Y})}}\right)_{\sun}}$} ratio is ${\rm [Ba/Eu]}=+1.13$, while the r-process-only ratio is ${\rm [Ba/Eu]}=-0.69$. In general, the europium and barium ratios indicate that HD\,140283 has a higher fraction of r-process elements than s-process elements \citep{Lambert2002,Mello2012}. In \citet{Gallagher2010} we set a lower limit for the [Ba/Eu] ratio, ${\rm [Ba/Eu]}>-0.66$, as limited quality in the spectrum around the europium 4129\,\AA\ line and uncertainties concerning blends around the europium 4205\,\AA\ line did not allow us to make a definitive measurement. However, this result suggests that both an r- and s-process regime are possible, considering the limits set in \citet{Arlandini1999}.

Most results agree well with currently accepted theory concerning the origin of the heavy elements in the early Galaxy. It was first proposed by \citet{Truran1981}, who inferred that the r-process should be dominant over the s-process at early times, explaining the observations and analysis of several metal-poor stars \citep{Spite1978}. This theory has subsequently shown to agree nicely with observational constraints on the s-process as a function of metallicity \citep{Francois2007}.

Several studies have examined the isotopic fractions of barium in HD\,140283 \citep{Magain1995,Lambert2002,Collet2009,Gallagher2010,Gallagher2012}, by measuring the asymmetry of the 4554\,\AA\ resonance line profile attributed to the odd isotopes of barium. The hyperfine structure (hfs) varies with the ratio, $\fodd$, which in turn reflects the nucleosynthetic origin of the isotopes\footnote{$\fodd=\left[N\left(\element[][135]{Ba}\right)+N\left(\element[][137]{Ba}\right)\right]/N\left(\element{Ba}\right)$}. For a 100\% r-process mixture, $\fodd=0.46$ and for a 100\% s-process mixture, $\fodd=0.11$ \citep{Arlandini1999}. The observational studies found a wide range of r- and s-process ratios: \citet{Magain1995} and \citet{Gallagher2010} found a high s-process fraction with almost no r-process material present, which conflicts with the \citeauthor{Truran1981} postulate. The \citet{Lambert2002} $\fodd$ value is intermediate between an r- and s-process solar system mix. The earlier studies used 1D stellar atmospheres, but later 3D atmospheres capable of studying the impact of velocity fields on line profile asymmetries were adopted \citep{Cayrel2007}. \citet{Collet2009} show differences between their 1D and 3D results for barium. From their 1D measurement, HD\,140283 shows an r-process signature but their 3D analysis drives the isotope fraction down to an s-process value, $\fodd=0.15\pm0.12$. One would infer from such results that this particular method of attaining the r- and s-process mixture is sensitive to the atmospheric modelling process.

Other methods of determining a star's barium isotope ratio have also been employed. \citet{Mashonkina1999} use non-local thermodynamic equilibrium (NLTE) model atmospheres and spectral synthesis to measure barium abundances. It is important that the method uses a NLTE approach, as very accurate abundance measurements are required, which a local thermodynamic equilibrium (LTE) technique cannot replicate to a high enough degree. They measure the abundance in high excitation barium lines (i.e. those other than the barium 4554 and 4934\,\AA\ resonance lines) and then force the two resonance lines to have the same abundance by increasing or decreasing the odd isotope fraction, while maintaining the same abundance in the subordinate lines. The ratio of the odd-to-even isotopes is then calculated and the r- and s-process ratio can be inferred. This method finds a good agreement with the [Ba/Eu] abundance determinations for the same star. Unfortunately, determining the $\fodd$ ratio in this fashion requires that barium has a high enough abundance that the weaker subordinate lines are strong enough to be distinguished from the continuum and indeed the noise in the spectrum. Therefore, using such a method to analyse HD\,140283 is not possible as the resonance and subordinate lines are very weak: the 4554\,\AA\ line has an equivalent width $W=20.1\,{\rm m\AA}$; the 4934\,\AA\ line's equivalent width is $13.6\,{\rm m\AA}$; the subordinate lines have an equivalent width range $4<W\,{\rm(m\AA)<13}$.

In \citet{Gallagher2010} (henceforth \po) we examined the barium isotopic fraction in detail by measuring $\fodd$ using the line profile asymmetry technique. In order to do this, it was necessary to analyse a number of iron lines so that the star's macroturbulent broadening could be determined, which is a necessary ``fudge factor'' that is included when working with 1D LTE spectral synthesis codes, since they do not calculate non-thermal motions. Iron was chosen over other elements as there are numerous lines to choose and it is free from hfs. (In \citet{Gallagher2012} we also examined calcium lines for this purpose, which agreed to within $1\sigma$ with the iron line determination.) To calculate the star's macroturbulence, 92 weak ($10\leq W\,({\rm m\AA})\leq50$) and apparently unblended iron lines were selected across the 4110--5650\,\AA\ wavelength range. Each line was modelled using \kurucz\ model atmospheres (\href{http://kurucz.harvard.edu/grids.html}{\tt http://kurucz.harvard.edu/grids.html}), with parameters $\Teff/\logg/{\rm [Fe/H]}=5750\,{\rm K}/3.7/-2.5$, and synthesised using the \atlas\ \citep{Cottrell1978} 1D LTE spectral synthesis code. They were fit to the observed spectrum while varying the abundance, macroturbulence and wavelength zeropoint, using a $\chi^2$ code, which is discussed at length in \po\ and \citet{Perez2009}. Large asymmetries were found in the red wing of the iron lines' residuals, which we speculated were due to a fundamental limit of 1D spectral modelling: the assumption of symmetric line broadening, and the neglect of 3D hydrodynamic motions of the gas. This was to be expected as it is well established that high resolution, high signal-to-noise spectra reveal asymmetric line profiles \citep{Gray1980}, even in stars with weak lines (like HD\,140283), that cannot be replicated by the classical 1D LTE approach to modelling a star's spectrum. In \po\ we speculated but did not attempt to confirm that the asymmetries encountered might be better replicated by employing the latest 3D time-dependent model atmospheres and spectral synthesis codes. In addtion, it was questioned whether a 3D treatment of the barium 4554\,\AA\ line would make any improvement to the isotope analysis. However, \citet{Collet2009} analysed the barium isotope fraction in both 1D and 3D. They found that the higher r-process fraction found under 1D, $\fodd=0.33\pm0.13$, was severely reduced under 3D to $\fodd=0.15\pm0.12$, close to the solar s-process-only ratio. This suggests that a 3D treatment of the barium line increases the inferred relative abundance of s-process-only isotopes and, hence reduces the r-process isotopes. 

In the present work we pursue the issue of line asymmetry by refitting the same set of iron lines used in \po\ and re-examine the barium isotope ratio, using the 3D LTE spectral synthesis code\footnote{\href{http://www.aip.de/Members/msteffen/linfor3d/files/linfor_3D_manual.pdf}{Link to the Linfor3D manual}} \linfor\ \citep{Steffen2010}, which employs time-dependent 3D model atmospheres produced with \cobold\ \citep{Freytag2012}. Elemental abundances are computed as a result of our methods, however, the discrepancies between classical and the 3D methodologies described in this paper have encouraged us to be mindful of these values.

This work is organised as follows: \S\ref{sec:observations} details the observations and reduction of the observed HD\,140283 spectrum; \S\ref{sec:modelsetup} and \S\ref{sec:modelanalysis} describe the details of the 3D hydrodynamic modelling process and examines various properties of the model atmospheres, respectively; \S\ref{sec:modellingiron} discusses the analysis of the iron lines in detail; \S\ref{sec:modellingbarium} presents the newly calculated $\fodd$ value for barium; and finally \S\ref{sec:conclusion} summarises our results.

\section{Observations}
\label{sec:observations}
The data are unchanged from \po: the stellar and thorium-argon calibration spectra were obtained during the commissioning of the High Dispersion Spectrograph (HDS) mounted on the Subaru Telescope. The stellar spectrum represents the sum of 13 exposures, taken over two nights (22/07/2001 and 29/07/2001), with a total exposure time of 82 minutes. This gives a $S/N=1100$ per 12\,m\AA\ wide pixel around 4500\,\AA, as measured from the scatter in the continuum of the reduced spectrum. The resultant stellar spectrum has a wavelength range 4110--5460\,\AA\ and 5520--6860\,\AA. The typical resolution as measured from thorium-argon lines is $R\equiv\lambda/\Delta\lambda = 95\,000$. The spectrum was reduced using the thorium-argon spectrum to wavelength calibrate the stellar spectrum, with typical RMS errors of 1.5\,m\AA\ \citep{Aoki2004}.

\begin{table*}[!t]
\begin{center}
\caption{The atmosphere parameters used in the present analysis. Temperatures are given to the nearest Kelvin. The Average $\Teff$ is the average of 20 retained snapshots, while the RMS errors of the 20 snapshots are listed to illustrate typical scatter in effective temperature between snapshots. $x,y,z$ represent the axes of the simulated box, where $z$ represents the vertical dimension. The number of opacity bins used in each model is given, and the computed time between the first and last selected snapshots is listed in the column labelled $\Delta \ {\rm time}$.}
\begin{tabular}{l c c c c r c c c r}
\hline\hline
Model name & Desired & $\logg$ & ${\rm [Fe/H]}$ & Average & ${\rm RMS}$ & Spatial resolution & Geometrical box & Opacity & $\Delta$ time \\
& $\Teff\,({\rm K})$ &&& $\Teff\,({\rm K})$ & ($\Teff$) & $[x,y,z]$ & size [$x,y,z$] (${\rm Mm}$)  & bins & (hours) \\
\hline
d3t57g37mm20n02 & $5750$ & $3.7$ & $-2.0$ & $5773$ & $9$ & $280\times280\times300$ & $36.6\times36.6\times22.2$ & $14$ & 5.3 \\
d3t57g37mm30n02 & $5750$ & $3.7$ & $-3.0$ & $5786$ & $10$ & $280\times280\times300$ & $36.6\times36.6\times22.2$ & $14$ & 10.6 \\
\vspace{-0.9em}\\
\hdashline
\vspace{-0.8em}\\
d3t57g37mm20n01 & $5750$ & $3.7$ & $-2.0$ & $5777$ & $11$ & $140\times140\times150$ & $36.7\times36.7\times22.2$ & $6$ & 26.4 \\
d3t57g37mm30n01 & $5750$ & $3.7$ & $-3.0$ & $5791$ & $11$ & $140\times140\times150$ & $36.7\times36.7\times22.2$ & $6$ & 26.4 \\
d3t55g35mm20n01 & $5500$ & $3.5$ & $-2.0$ & $5502$ & $14$ & $140\times140\times150$ & $49.0\times49.0\times35.9$ & $6$ & 13.0 \\
d3t55g35mm30n01 & $5500$ & $3.5$ & $-3.0$ & $5536$ & $12$ & $140\times140\times150$ & $49.0\times49.0\times35.9$ & $6$ & 56.0 \\
d3t55g40mm20n01 & $5500$ & $4.0$ & $-2.0$ & $5473$ & $7$ & $140\times140\times150$ & $20.1\times20.1\times10.6$ & $6$ & 9.4 \\
d3t55g40mm30n01 & $5500$ & $4.0$ & $-3.0$ & $5475$ & $6$ & $140\times140\times150$ & $20.3\times20.3\times10.6$ & $6$ & 16.9 \\
d3t59g35mm20n01 & $5900$ & $3.5$ & $-2.0$ & $5862$ & $14$ & $140\times140\times150$ & $59.3\times59.3\times38.7$ & $6$ & 31.1 \\
d3t59g35mm30n01 & $5900$ & $3.5$ & $-3.0$ & $5873$ & $14$ & $140\times140\times150$ & $59.3\times59.3\times38.7$ & $6$ & 31.1 \\
d3t59g40mm20n01 & $5900$ & $4.0$ & $-2.0$ & $5857$ & $7$ & $140\times140\times150$ & $25.8\times25.8\times12.5$ & $6$ & 8.3 \\
d3t59g40mm30n01 & $5900$ & $4.0$ & $-3.0$ & $5851$ & $7$ & $140\times140\times150$ & $25.8\times25.8\times12.5$ & $6$ & 8.5 \\
\hline
\label{tab:atmtemp}
\end{tabular}
\end{center}
\end{table*}

\section{Modelling HD\,140283}
\label{sec:modelsetup}

The atmosphere parameters used for modelling HD\,140283 were set to mirror those in the 1D LTE study carried out in \po, as closely as possible. The parameters that were used to construct the 1D LTE atmospheres were $\Teff=5750\,{\rm K}$, $\logg=3.7$, ${\rm [Fe/H]}=-2.5$. The same $\logg$ value was selected for the 3D atmospheres. Unlike a 1D model atmosphere, in a 3D model atmosphere the effective temperature is not a control parameter. Instead, values for entropy are set for the material flowing into the computational box through the open lower boundary; a higher entropy here will lead to a larger flux, and hence higher effective temperature \citep{Caffau2009}. Suitable values of entropy were selected so that comparable effective temperatures were seen for both the 1D and 3D model atmospheres. Opacity tables for ${\rm [Fe/H]}=-2.5$ were not available at the time of this analysis. As a result, we computed pairs of model atmospheres, one for ${\rm [Fe/H]}=-2.0$ and the other for ${\rm [Fe/H]}=-3.0$. 

Table~\ref{tab:atmtemp} tabulates the twelve parameter sets of the \cobold\ \citep{Freytag2012} 3D model atmospheres we investigated: six of them have been computed for ${\rm [Fe/H]}=-2.0$ and the other six for ${\rm [Fe/H]}=-3.0$. The first four parameter sets that best describe HD\,140283 were computed for this work. The other eight with different temperature and $\logg$ values were constructed as part of the ``Cosmological Impact of the First STars'' (CIFIST) collaboration \citep{Ludwig2009}, and are used to determine the sensitivity to derived parameters. Other than when stated, only the two high resolution atmospheres are used.

Three types of atmosphere were computed for each model parameter set: a fully 3D model atmosphere computed with \cobold\ (\S\ref{sec:cobold}), an equivalent average 3D model atmosphere, referred to as \tda\ (see \S\ref{sec:avcobold}), and a 1D hydrostatic model atmosphere computed by the Lagrangian 1D hydrodynamical (\od) model atmosphere code. The reason for computing \od\ models is noted in \S\ref{sec:lhdmodel}. 

All these models employ opacities based on the MARCS stellar atmosphere package \citep{Gustafsson2008}. In \S\ref{sec:modelanalysis} we compare the 3D and \tda\ atmospheres with 1D atmospheres generated using the \od\ package, while in \S\ref{sec:modellingiron}--\ref{sec:modellingbarium} we compare the resultant 3D synthesis with 1D synthesis computed using the 1D LTE spectral synthesis code \atlas\footnote{The \citeauthor{Cottrell1978} code called {\tiny ATLAS} should not be confused with the model atmospheres and code of the same name by Kurucz.} \citep{Cottrell1978}, which we employ with 1D LTE Kurucz model atmospheres as in \po. It is expected that differences in resultant 1D syntheses from both atmospheres would be negligible, as it has been shown before that differences between the Kurucz and MARCS model atmospheres are immaterial \citep{Bonifacio2009}. Indeed, results from comparisons of both 1D techniques show that differences in line strengths produced from either technique for a given abundance are small for all iron lines analysed in the sample -- less than $0.4\,{\rm m}\AA$ on average, with a line-to-line scatter of $0.74\,{\rm m}\AA$. These differences suggest that synthesis using either 1D method would produce lines with equal abundances for a given spectral feature.

\subsection{The 3D \cobold\ atmospheres}
\label{sec:cobold}

The 3D models that best describe HD\,140283 (i.e. $\Teff=5750\,{\rm K}$, $\logg=3.7$, ${\rm [Fe/H]}=-2.0$ \& $-3.0$) were computed for high and low spatial resolutions, using the ``box-in-a-star'' set-up for a small proportion of the stellar atmosphere. Both cover the same physical region of a star (covering optical depths $10^{-6.5}$ to $10^{5}$), but the high resolution models have twice as many voxels in each Cartesian axis (i.e. eight times as many voxels overall) and opacities are sampled at more than twice the number of frequency intervals. The high resolution models were only computed for the best prescription of HD\,140283 because of the length of time required to compute them. All models consist of a series of 20 temporal structures, which we refer to as snapshots.

The geometrical size of the computational box is approximately scaled according to the model's temperature, $\logg$, and metallicity by scaling the box size according to the resultant pressure scale height at the surface relative to the solar model geometrical size, $5.6\times5.6\times2.3\,{\rm Mm}$. However, metallicity effects have a small impact on this scaling factor and are ignored so that models that share the same temperature and $\logg$ but differ in metallicity can be compared without the need to rebin the computational box \citep{Caffau2009}. Effects related to stellar sphericity are neglected as a consequence of the size of the computational box, relative to the size of the star (radius of HD\,140283 $\sim2\,\rsun$\footnote{Based upon $\logg=3.7$ and $M_{*}=0.8\,\Msol$}). The computational box corresponds to $\sim0.01\%$ of the surface of one hemisphere of HD\,140283.

\subsection{The average \tda\ \cobold\ atmospheres}
\label{sec:avcobold}
Average \tda\ model atmospheres are produced by spatially averaging the thermal structure of the 3D computational box over surfaces of equal Rosseland optical depth ($\tau_{\rm ROSS}$). This was done twenty times, once for each of the 20 selected snapshots that make up every atmosphere tabulated in Table~\ref{tab:atmtemp}. Details of the procedure that produces \cobold\ atmospheres, like the ones used in this study, can be found in \citet{Ludwig2012,Kucinskas2013}.

\begin{figure*}[!ht]
\begin{center}
	\resizebox{\hsize}{!}{\includegraphics{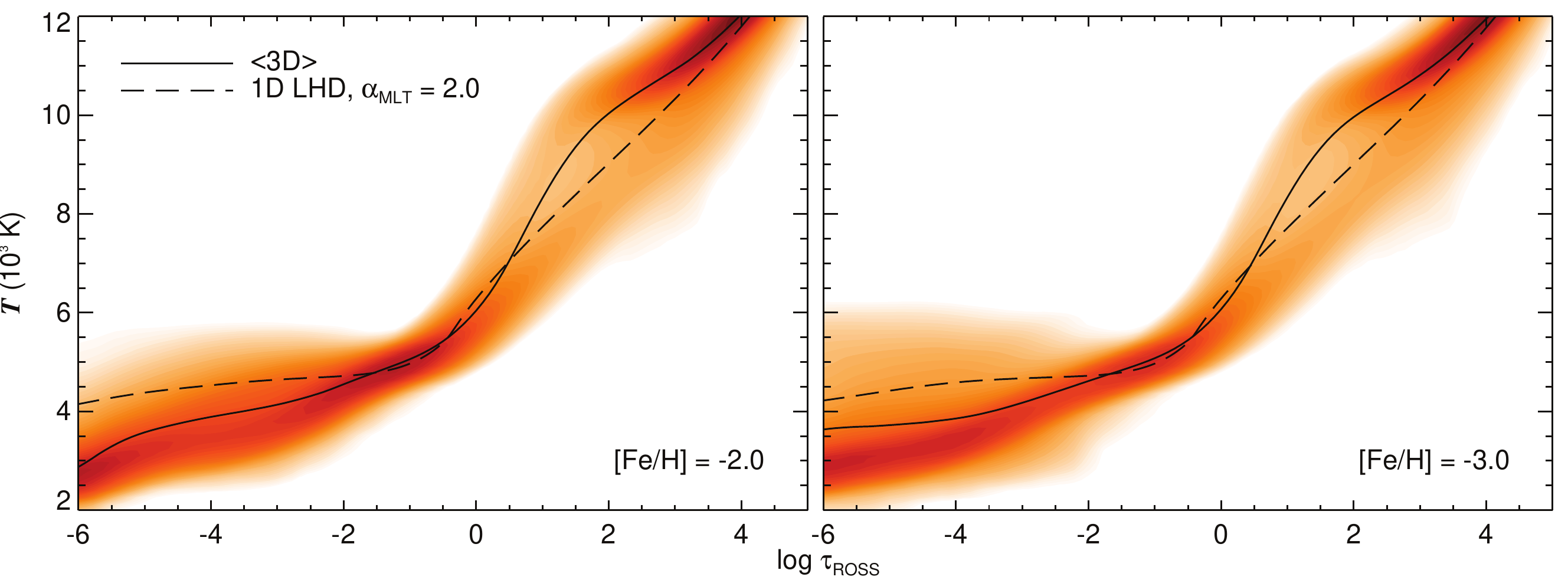}}
	\caption{Temperature structures of: the full 3D atmospheres based on $280\times280\times300$ grids points (orange density plot); the average \tda\ atmosphere (solid black line); and the equivalent 1D \od\ temperature structure (dashed black line) for the ${\rm [Fe/H]}=-2.0$ (left) and ${\rm [Fe/H]}=-3.0$ (right) high resolution atmospheres, for a single snapshot in time. Darker regions in the density plot represent areas in the 3D atmospheres with densely populated temperature regions for a given optical depth. Large temperature deviations exist between the \tda\ and 1D atmospheres in both the deep layers ($\ltaur>1.0$) of the star and in line forming regions at $\ltaur<-2.0$, while in layers where $T\sim\Teff$, the \tda\ and 1D temperature structures are quite similar.}
	\label{fig:3d1dts}
\end{center}
\end{figure*}

\subsection{The 1D model atmospheres}
\label{sec:lhdmodel}
Every 3D model atmosphere described in Table~\ref{tab:atmtemp} has a counterpart 1D model atmosphere, produced with the \od\ code. They were produced using the same atmospheric parameters, chemical compositions, opacities and equations of state as those used to compute the \cobold\ 3D and \tda\ model atmospheres. Convection is treated using the \citet{Mihalas1978} formulation of the mixing length theory (MLT). Three 1D atmospheres were produced for every parameter set-up described in Table~\ref{tab:atmtemp}, for $\alpha_{\rm MLT}=0.5$, $1.0$ and $2.0$. We discuss the effects that the choice in the MLT parameter has on the 1D atmosphere in \S\ref{sec:tempstructure}.

\section{Atmospheric model analysis}
\label{sec:modelanalysis}
In this section we detail notable deviations between the \cobold\ and equivalent 1D \od\ model atmospheres. We also discuss the impact that chemical composition has on certain parameters that line formation is most sensitive to. Explanations discussed here help to understand some properties that the 3D synthesis exhibit, relative to the 1D synthesis, which are discussed throughout \S\ref{sec:modellingiron}--\ref{sec:modellingbarium}. 

\subsection{Temperature structure}
\label{sec:tempstructure}

The temperature structure ($T,\tau$ relation) of a 3D simulation differs even along parallel lines of sight. This makes it difficult to compare it with a classical 1D model atmosphere. Using the \tda\ temperature structures (\S\ref{sec:avcobold}) makes the comparison between 1D and 3D simpler, as Fig.~\ref{fig:3d1dts} illustrates.

One can see notable differences in temperature ($\sim800\,{\rm K}$) between the \tda\ and 1D atmospheres at $\ltaur<-4.0$. While the 1D model temperature profile begins to plateau, the \tda\ models continue to cool outwards. \citet{Asplund1999} and \citet{Asplund2001} attribute this cooling effect to the inefficient radiative heating of the outermost layers of low metallicity model atmospheres relative to cooling by overshooting.

\begin{figure}[!t]
\begin{center}
	\resizebox{\hsize}{!}{\includegraphics{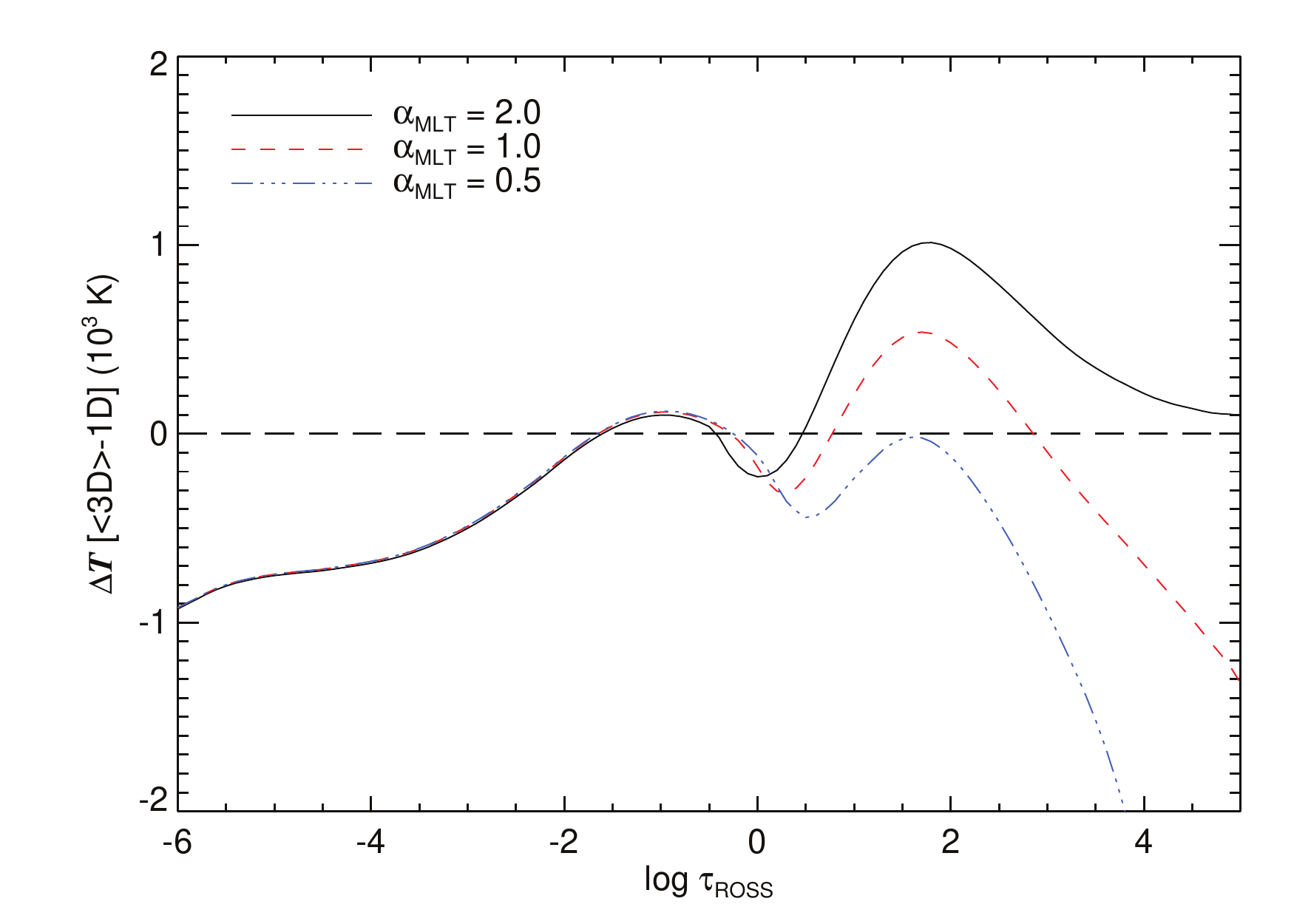}}
	\caption{The temperature structure differences between the \tda\ and 1D model atmospheres for three values of $\alpha_{\rm MLT}$. The $(T,\tau)$ profiles for the ${\rm [Fe/H]}=-2.0$ and $-3.0$ \tda\ atmospheres have been averaged to produce the \tda\ reference used in generating this plot. The outer atmosphere is not sensitive to values of $\alpha_{\rm MLT}$, while the deeper layers of the atmosphere are extremely sensitive to this parameter.}
		\label{fig:d3d1tdeviation}
\end{center}
\end{figure}

Similarly, notable differences in temperature appear deeper in the star at $\ltaur>1$. This can be controlled by the mixing length parameter ($\alpha_{\rm MLT}$) selected for the 1D model. In Fig.~\ref{fig:3d1dts}, $\alpha_{\rm MLT}=2.0$. Selection of different values for $\alpha_{\rm MLT}$ in the 1D model are still unable to adequately reproduce the temperature structure seen in the \tda\ model \citep{Ludwig2012}; Fig.~\ref{fig:d3d1tdeviation}. However, as lines do not form in these regions, it is of little consequence to the work presented here. We have illustrated the effects that $\alpha_{\rm MLT}$ has on the 1D temperature structure in Fig.~\ref{fig:d3d1tdeviation} relative to an interpolated \tda\ model atmosphere for ${\rm [Fe/H]}=-2.5$\footnote{All other plots presented in this paper are produced by interpolating the ${\rm [Fe/H]}=-2.0$ and $-3.0$ \tda\ ($T,\tau$) structures together unless otherwise stated.}. While the line forming region in the outer atmosphere is relatively unchanged for varying $\alpha_{\rm MLT}$, the deeper regions of the atmosphere are heavily dependent on this value.

The large temperature differences between the \tda\ and 1D model atmospheres seen in the outer atmosphere could lead to large deviations in line strengths of absorption lines that form there, which will lead to large abundance corrections between the 3D and 1D models. As the grid points in the outermost layer of a fully 3D model atmosphere do not all have the same physical characteristics, grid points in that layer correspond to a range of (low) optical depths. Inspection of the models shows that $84\%$ of the voxels in that layer have optical depths $\ltaur<-6.0$, indicating that model data presented (e.g. Fig.~\ref{fig:d3d1tdeviation}) at $\ltaur\geq-6.0$ should be little affected by numerical boundary effects. Deeper layers where absorption lines and the continuum form will clearly be unaffected.

\subsection{Ionisation fractions}
\label{sec:ionisation}

\begin{figure}[!t]
\begin{center}
	\resizebox{\hsize}{!}{\includegraphics{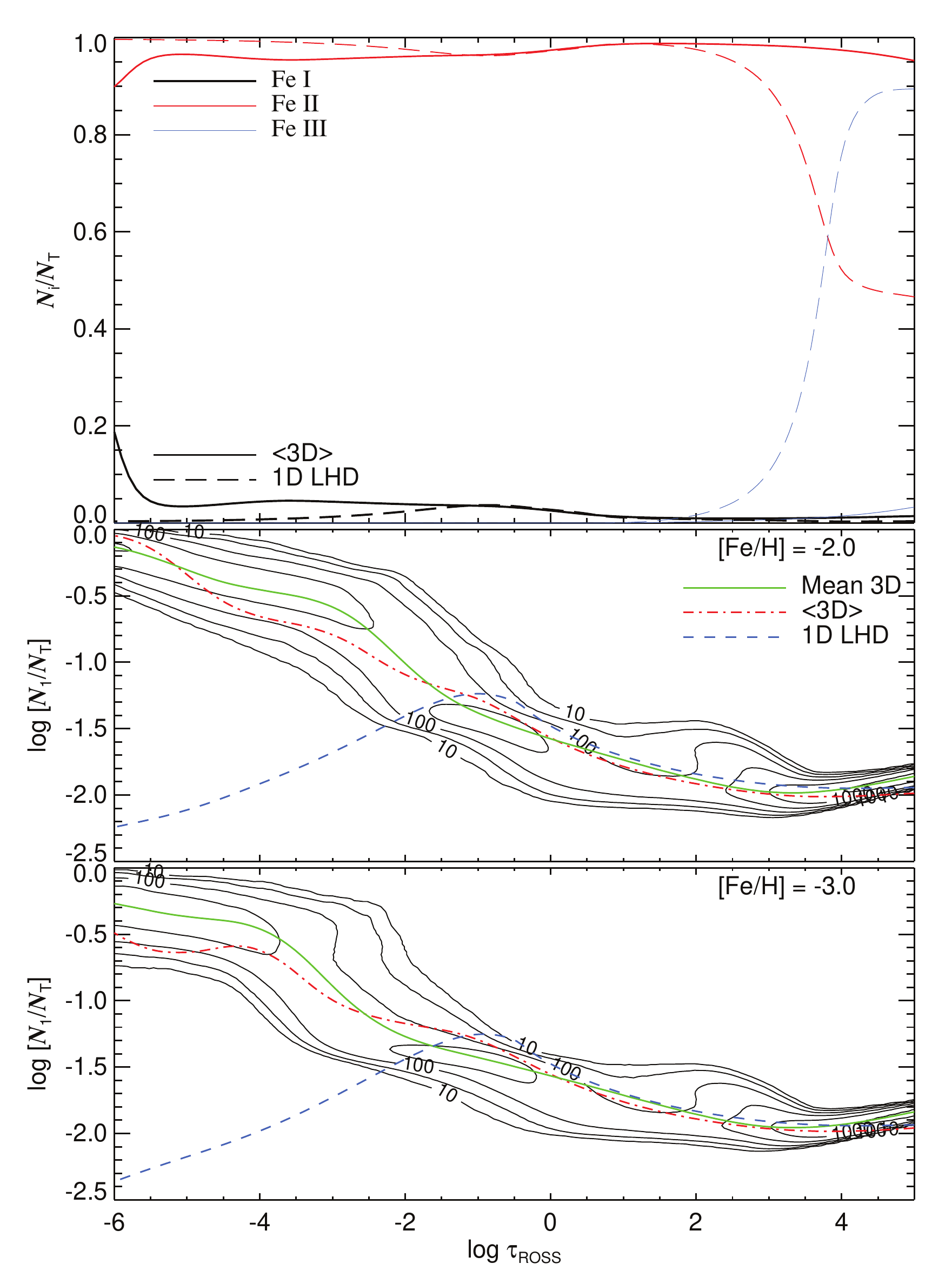}}
	\caption{(\textit{Top panel}): Comparisons of the \tda\ (solid lines) and 1D (dashed lines) atmosphere's ionisation fractions for the neutral and first two ionisation levels of iron. The ${\rm [Fe/H]}=-2.0$ and $-3.0$ \tda\ atmospheres have been interpolated to produce this plot. (\textit{Middle panel}): Ionisation fraction of \fei\ for the full 3D (contour) atmosphere for 10 snapshots of the ${\rm [Fe/H]} = -2.0$ atmosphere. Contour values represent non-normalised frequencies. Also included are the equivalent \tda\ (dash--dot), average 3D (solid line) and 1D \od\ (dashed line) ionisation fraction of \fei. (\textit{Bottom panel}): Same as middle panel for the ${\rm [Fe/H]} = -3.0$ atmosphere. All ionisation fractions presented in the figure were calculated under the assumption of LTE.}
	\label{fig:ionisation}
\end{center}
\end{figure}

The lower temperatures in the outer regions of 3D model atmospheres, relative to the equivalent 1D model atmospheres, will have an effect on the ratio of neutral to singly ionised iron absorbers. It is well understood from the Saha equation that in an atmosphere such as those used here ($\Teff=5750$, $\logg=3.7$), \feii\ is the dominant ionisation stage for iron. Saha's equation relates the first ionised-neutral number density ratio, $\frac{N_{1}}{N_{0}}$, and the electron pressure, $P_{\rm e}$, to terms depending essentially only on the temperature, $T$, the identity of the atomic species, $Z$, via its ionisation potential, $I_Z$, and partition functions, e.g. $\frac{N_{1}}{N_{0}}P_{\rm e}=f(T,Z)$. The temperature is almost uniform in the outer layers ($\ltaur<-1.5$) of the 1D model, and as a result $\frac{N_{1}}{N_{0}}P_{\rm e}$ is to first order constant there. At lower optical depths, the gas pressure falls as the density falls, and with it the source of electrons and hence electron pressure. As $P_{\rm e}$ falls, $\frac{N_{1}}{N_{0}}$ must rise, but as $\frac{N_{1}}{N_{1}+N_{0}}$ is already nearly $100\%$ and hence $N_{1}$ has little room for adjustment, instead $N_{0}$ must fall. Hence, for $\ltaur<-1.5$, $N_{0}$ falls at smaller $\ltaur$ in the roughly isothermal 1D atmosphere (Fig.~\ref{fig:ionisation}).

Fig.~\ref{fig:ionisation} illustrates that the 3D and 1D iron fractions are almost identical between $-1.0\leq\ltaur\leq2.0$ (the lower two panels show that there are slight deviations), but as the outer 3D atmosphere cools more efficiently, $\frac{N_{1}}{N_{0}}P_{\rm e}=f(T,\tau)$ is no longer constant, and falls. At lower optical depths, we see from the lower two panels of Fig.~\ref{fig:ionisation} that the neutral fraction of iron is steadily increasing over the range $-5\leq\ltaur\leq-1$. Compared with the 1D model, the neutral fraction is several times higher. The singly ionised fraction is lower to compensate, but only by a very small factor ($<5\%$). This has a noticeable secondary impact on the abundances derived for \fei\ lines in 3D, but not of course for \feii. One effect (which we will see in \S\ref{sec:feformation}) is that the \fei\ lines in 3D model atmospheres form much further out than in 1D atmospheres, but \feii\ lines are less affected.

\section{Modelling iron lines}
\label{sec:modellingiron}

In this section we detail the behaviours shown by the 3D synthesis, relative to the 1D synthesis presented in \po. 

\subsection{Synthesis with \linfor}
\label{sec:linforsynthesis}
Synthetic spectral line profiles were computed using the 3D spectral synthesis code \linfor\ for 91 of the 92 lines analysed in \po. The Fe\,I 5107.45\,\AA\ line was dropped from the 3D analysis owing to a blend with another iron line, redward of the line centroid. 

Synthesising a 3D spectrum is very time consuming, therefore it is impractical to compute a spectrum for the entire observed wavelength range. For a detailed review of the computational considerations for synthesis with Linfor3D, see \citet{Bonifacio2013}. As such, the computations were limited to a small wavelength range of $\pm0.3$\,\AA\ around each line centroid. Computational times varied according to the spatial resolution of the \cobold\ atmospheres, the wavelength sampling (set to $5\times10^{-3}$\,\AA) and the wavelength range over which the synthesis was calculated.

At the time of this research, the opacities necessary for a 3D model with metallicity ${\rm [Fe/H]}=-2.5$ were not available. As such, every 3D synthetic profile used in the present analysis represents an interpolation of spectra computed separately for model atmospheres with ${\rm [Fe/H]}=-2.0$ \& $-3.0$. The abundance is altered by changing the $\loggf$ of each line. For each atmosphere, every line was synthesised for 11 abundances (i.e. 11 $\loggf$ values) over a range of 0.4 dex, in steps of 0.04 dex, for line strengths that closely matched the observed line's equivalent width. As such, the $\loggf$ values used to produce lines with the ${\rm [Fe/H]}=-3.0$ model atmosphere were 1 dex larger than those used to construct lines with the ${\rm [Fe/H]}=-2.0$ model atmosphere so that the resulting synthesis from both atmospheres would have approximately the same equivalent width. 

In addition, the line profiles were convolved with a Gaussian of $FWHM=3.31\,\kms$, which is used to represent the instrumental broadening of the observed spectrum, as ascertained in \po.

\subsection{External broadening determination}
\label{sec:vsini}
Analogous to a 1D LTE investigation, it is important that all external broadening parameters are adequately resolved so that the isotope ratio determination is not affected. From \citet{Lambert2002} and the work conducted in \po, it is clear that barium isotope ratios determined in the manner prescribed are extremely sensitive to external broadening effects, as $\delta\fodd/\delta v_{FWHM}$ was found to be $-0.51$ and $-0.7\,(\kms)^{-1}$, respectively.

We have re-analysed all 91 iron lines to determine the rotational speed, $\vsini$, of HD\,140283, and assume this to account for broadening not already modelled by the 3D atmospheres or the instrumental broadening. The iron line sample contains 81 \fei\ and 10 \feii\ lines. Similar to \po, each line was individually fit using a $\chi^2$ routine, described in \citet{Perez2009}. The code determines the best fit synthetic profile by $\chi^2$ minimisation\footnote{Using the definition $\chi^{2}=\Sigma\left[\left(O_{i}-M_{i}\right)^{2}/\sigma_{i}^{2}\right]$, where $O_{i}$, $M_{i}$ and $\sigma_{i}$ are the observed profile, model profile and error estimation in the observed profile, $\sigma=\left(S/N\right)^{-1}$, respectively.}, for varying wavelength shift, iron abundance ($\afe$) and, in this work, $\vsini$. The continuum normalisation is also calculated and included when determining the three free parameters for each line by measuring small regions of the continuum, either side of every iron line.

We model the rotational profile using $\vsini$ profiles as implemented in \linfor\ \citep{Ludwig2007}, which are convolved to the synthetic spectra. In total, a grid of $121$ synthetic spectra were computed for every iron line in the sample, i.e. for $11$ values of $\vsini$ and $11$ values of $\afe$. $\vsini$ values were computed over the range $1.0\leq\vsini \ (\kms)\leq6.0$ with $\Delta\vsini=0.5\,\kms$. The iron abundances in the synthetic grid covered a small abundance range of 0.4\,dex in steps $\Delta\afe=0.04$\,dex. Therefore the $\afe$ grid was unique to each line as the differences between iron abundance across the line sample is larger than the range of $\afe$ values that needed to be covered by the synthesis for an individual line. 

\subsubsection{Result}
\label{sec:vsini_result}
\begin{figure}[!t]
\begin{center}
	\resizebox{\hsize}{!}{\includegraphics{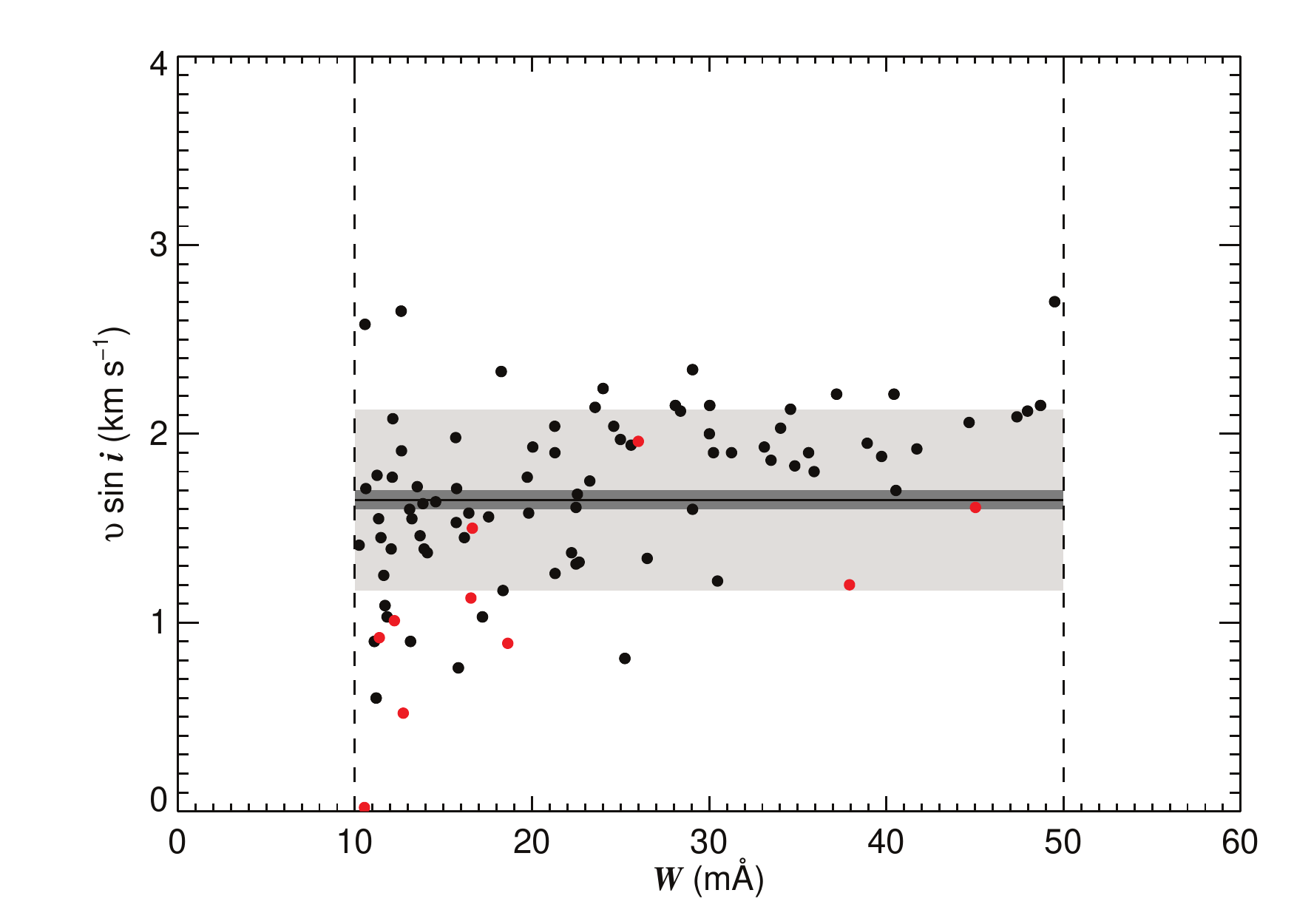}}
	\caption{The behaviour of the measured $\vsini$ from the interpolated spectrum, relative to the line strength for the 81 \fei\ (black dots) and 10 \feii\ (red dots) lines. The lightest shaded region shows the scatter relative to the mean from the sample ($\sigma$); the darker shaded region the standard error ($0.05$); and the darkest line represents the mean value ($1.65\,\kms$) used in the analysis.}
		\label{fig:vsiniEW}
\end{center}
\end{figure}

It was found that $\vsini=1.65\pm0.05\,\kms$, where the error represents the random error associated with the line-to-line scatter. In Fig.~\ref{fig:vsiniEW} we have plotted the relationship between $\vsini$ and line strength. It is shown that the scatter in $\vsini$ increases at smaller line strength, demonstrating the effect of finite $S/N$. It also shows that the dispersion is asymmetric about the mean $\vsini$ value (represented by the dark line) and the scatter, relative to the mean (given by the lightly shaded region). 

Further testing of the \fei\ and \feii\ lines revealed two contrasting values for $\vsini$: the average $\vsini$ value for the \fei\ lines was $1.72\pm0.05\,\kms$, while the average $\vsini$ value for the \feii\ lines was $1.08\pm0.17\,\kms$, where errors represent the standard error associated with line-to-line scatter in each case. It is possible that this behaviour is indicative of NLTE influences. Similar behaviours were shown in \fei\ lines and lithium under LTE in \citet{Lind2013a} and in lithium in \citet{Steffen2012}. When NLTE effects were considered in the line synthesis, \fei\ $\vsini$ values were reduced towards \feii\ $\vsini$ values; the NLTE profiles were broader than the LTE profiles.

The corresponding mean iron abundance is $\afe=4.78\pm0.05\pm0.02$, where the errors represent the random error associated with line-to-line scatter added in quadrature to the uncertainties associated with temperature and gravity determinations of HD\,140283 (\S\ref{sec:atm_errors}) and the systematic error associated with choice in model atmosphere, but as we discuss in \S\ref{sec:fecorrections}, there are considerable systematic features of the iron abundance which need further elaboration.

\subsubsection{Individual atmosphere results}
\label{sec:atm_effects}
The synthetic iron line spectra produced from the ${\rm [Fe/H]}=-2.0$ \& $-3.0$ model atmospheres (which are both assembled by averaging over the 20 snapshots that make up each atmosphere) were also fit to the observed spectrum. The ${\rm [Fe/H]}=-2.0$ model atmosphere was found to have an iron abundance of $4.76\pm0.02\,{\rm dex}$ and $\vsini$ was found to be $1.89\pm0.05\,\kms$. The ${\rm [Fe/H]}=-3.0$ was found to have an iron abundance of $4.79\pm0.02\,{\rm dex}$ and $\vsini$ was found to be $1.57\pm0.05\,\kms$. No noteworthy improvements were seen between the fits for either atmosphere and no other noticeable effects were detected.

\subsection{3D--1D abundance differences}
\label{sec:fecorrections}

\begin{figure}[!t]
\begin{center}
	\resizebox{\hsize}{!}{\includegraphics{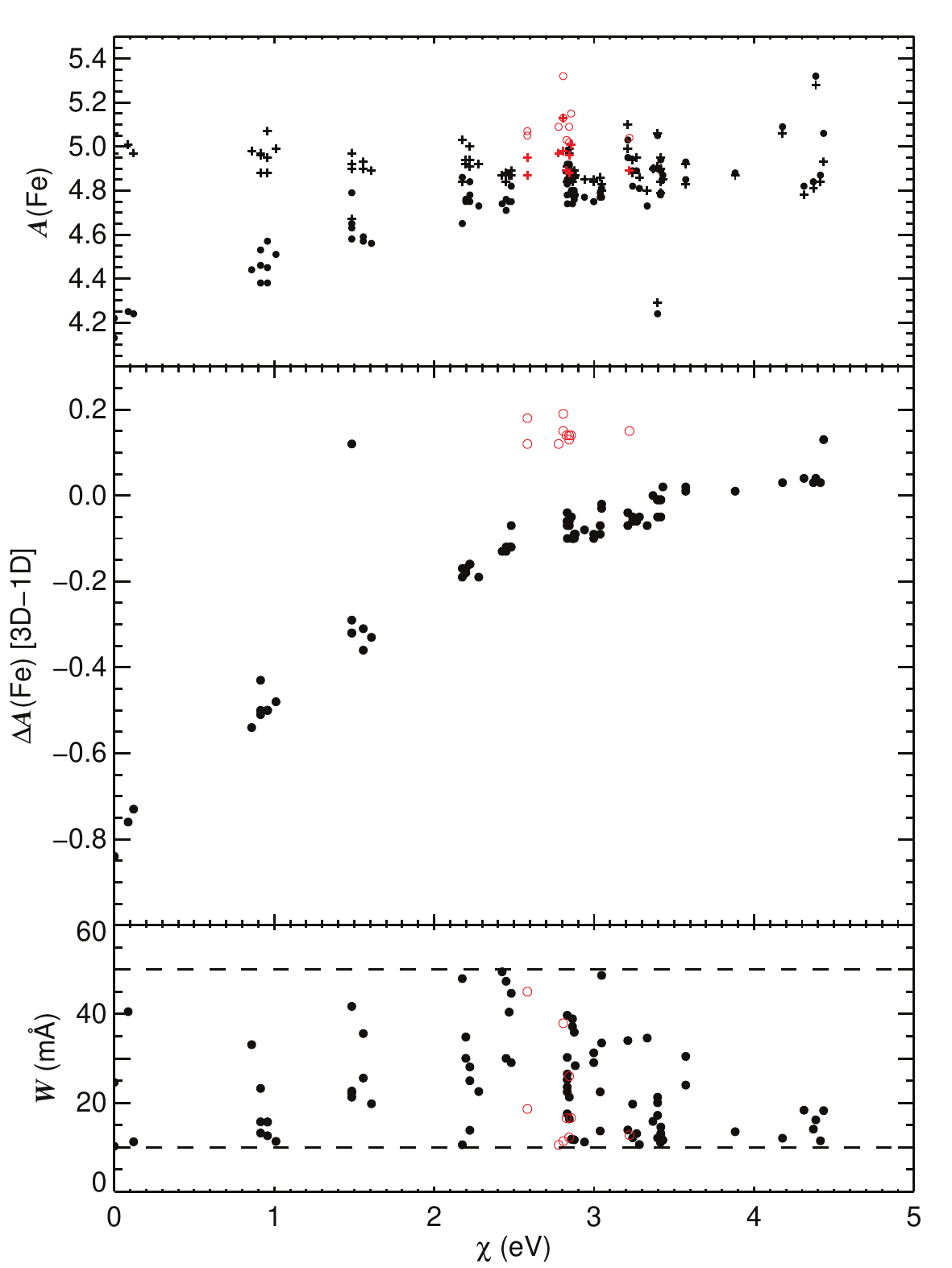}}
	\caption{\textit{Top panel}: The 3D (filled circles) and 1D (crosses) abundances for the 91 iron lines analysed under LTE. Red open circles and red crosses represent the \feii\ lines for the 3D and 1D synthesis, respectively. While the 1D abundances remain fairly consistent for varying excitation potential, the 3D results show a reduction in abundance at low excitation potentials. \textit{Middle panel}: The apparent excitation potential dependence shown by the \fei\ lines (filled circles), emphasised by plotting the difference between 3D and 1D abundances. The \feii\ lines (red open circles) are not affected by this trend. The clear \fei\ outlier seen at $\sim 1.5\,{\rm eV}$ is caused by a poor fit to the iron line under the 1D paradigm from \po. \textit{Bottom panel}: The scatter in line strength with excitation potential for the \fei\ (filled circles) and \feii\ (red open circles). This also shows that there is no line strength dependence on excitation potential and thus the abundance difference in the middle panel is purely an effect of excitation potential.}
	\label{fig:chidependence}
\end{center}
\end{figure}

It has been established in previous works that deviations in line abundance exist between 3D and 1D analyses \citep{Asplund1999,Asplund2001,Frebel2008,Kucinskas2013}. The analysis done here likewise found abundance differences in the 3D analysis, relative to the 1D analysis in \po. Initially we only examine the 81 \fei\ lines. In \S\ref{sec:feivsfeii}, we discuss the 10 \feii\ lines. 

\subsubsection{Excitation potential}
\label{sec:excitation_potential}
Fig.~\ref{fig:chidependence} shows that there is a clear \fei\ abundance dependence on excitation potential. From the top panel it is shown that while the \fei\ lines synthesised under 1D show no trend with excitation potential, the \fei\ lines synthesised under 3D show do a definite trend. In the second panel we show the difference between the 3D and 1D abundances. The trend of the \fei\ abundances with excitation potential is extremely well defined; lines with very low excitation potential require large reductions in abundance under 3D to maintain the same equivalent width as the observed profile. In the bottom panel we have shown that there is no major correlation between excitation potential and line strength, so the trend in the middle panel is indeed an excitation potential effect. This effect has been seen in other published works, such as \citet{Collet2009a} who examine carbon, nitrogen, oxygen and iron abundances in the metal-poor giant HD\,122563. They speculate that said effect may be due to the differences in the 3D and 1D temperature structures, shortcomings in their model atmospheres and/ or the departures from LTE in the models they present. In the following two sections, we demonstrate that the effect is caused by differences in the thermal structure between the 1D and 3D models.

\begin{figure}[!t]
\begin{center}
	\resizebox{\hsize}{!}{\includegraphics{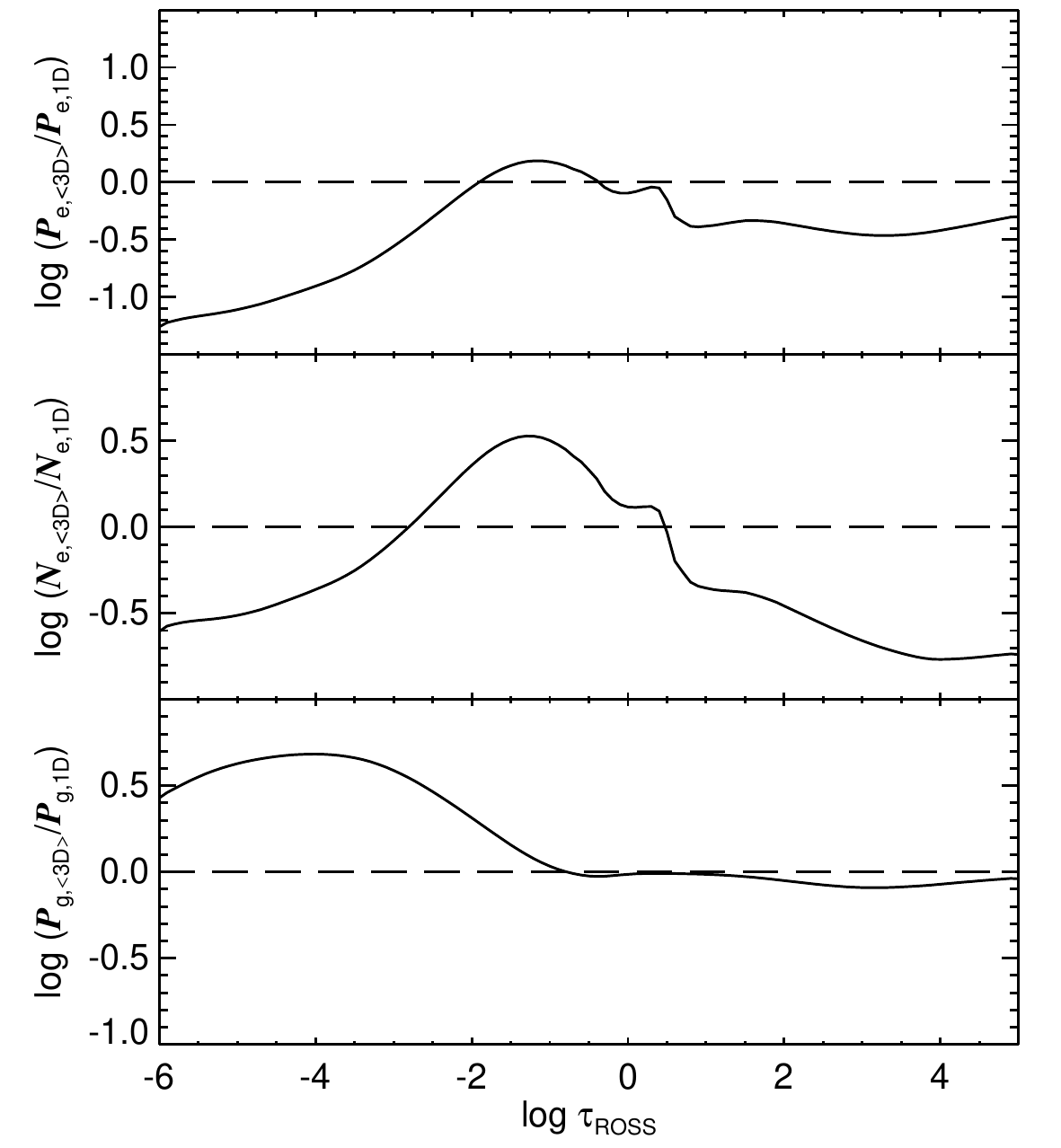}}
	\caption{The \tda\ model electron pressure (top panel), electron density (middle panel) and gas pressure (bottom panel), relative to the equivalent 1D model.}
	\label{fig:edensity}
\end{center}
\end{figure}

Figs~\ref{fig:3d1dts} \& \ref{fig:d3d1tdeviation} illustrate that at $\ltaur<-4$, temperature differences approach $800\,{\rm K}$. These differences lead to a dramatic decrease in the electron pressure in the 3D atmosphere, relative to the equivalent 1D atmosphere, as shown in the top panel of Fig.~\ref{fig:edensity}. At the outermost parts of the atmospheres ($\ltaur<-4.5$), the difference in electron pressure is over an order of magnitude. In the middle panel of Fig.~\ref{fig:edensity} we have plotted the corresponding relative electron densities of the two model atmospheres, calculated from the electron pressures, assuming $P_{\rm e}=N_{\rm e}kT$. The large decrease in the electron number density in the 3D model, relative to the 1D is related to the higher neutral fraction seen at $\ltaur<-3$ in Fig.~\ref{fig:ionisation}. They suggest that the 3D atmospheres have cooled sufficiently to allow electrons to recombine with the ionised iron, whereas temperatures are still too high in the equivalent 1D models. The bottom panel of Fig.~\ref{fig:edensity} illustrates that the relative behaviours of the electron pressures are not driven by a rapid decrease in the \tda\ gas pressures, relative to the 1D, as the \tda\ gas pressures remain higher than the equivalent 1D gas pressure in the outer atmosphere.

It is of course disconcerting that the abundances derived in 3D have a significant dependence on excitation potential. Similar trends were found when model atmospheres with higher and lower effective temperatures were selected from Table~\ref{tab:atmtemp}. \citet{Bergemann2012} show that this trend is removed when the analysis is conducted under a \tda\ NLTE paradigm. This would strongly suggest that this unwelcome finding is a feature of the assumption of LTE in the line synthesis, and not merely because of a poor choice in effective temperature or a feature of the 3D (and also \tda) models. This also implies that the systematic error associated with $\afe$ under LTE, presented in \S\ref{sec:vsini_result}, is a poor estimate of the true systematic error. The excitation potential was checked against $\vsini$ and no trend was found, implying that $\vsini$ remains unaffected by this effect. To explore this feature further, we investigated the formation depth of the lines.

\begin{table*}[!t]
\begin{center}
\caption{Various properties of the four selected lines of study. Columns $(2)$ and $(3)$ tabulate the resultant iron abundance found from the 3D analysis conducted in the present work, and the 1D analysis from {\scriptsize PAPER1}. The excitation potential and $\loggf$ values have also been tabulated. Column $(6)$ tabulates the observed equivalent widths of the four lines. Column $(7)$ tabulates the equivalent widths of the 3D synthesis, computed with the 3D abundances from column $(2)$ ($W_{\rm 3D,3D}$). Column $(8)$ tabulates the 1D synthesis computed with the 1D abundances tabulated in column $(3)$ ($W_{\rm 1D,1D}$). Column $(9)$ tabulates the 1D synthesis computed with the abundances from column 2 ($W_{\rm 1D,3D}$).}
\begin{tabular}{l c c c c c c c r}
\hline\hline
$\lambda$ (\AA) & $\afe_{\rm 3D}$ & $\afe_{\rm 1D}$ & $\chi$ & $\loggf$  & $W_{\rm obs}$ & $W_{\rm 3D,3D}$ & $W_{\rm 1D,1D}$ & $W_{\rm 1D,3D}$ \\
& & & (eV) & & & (m\AA) & (m\AA) & (m\AA) \\
$(1)$ & $(2)$ & $(3)$ & $(4)$ & $(5)$ & $(6)$ & $(7)$ & $(8)$ & $(9)$ \\
\hline
$4132.900$ & $4.92$ & $4.99$ & $2.845$ & $-1.005$ & $16.43$ & $15.62$ & $15.57$ & $14.03$ \\
$4143.410$ & $4.81$ & $4.83$ & $3.047$ & $-0.204$ & $33.49$ & $34.44$ & $34.86$ & $32.51$ \\
$4461.653$ & $4.25$ & $5.01$ & $0.087$ & $-3.194$ & $40.54$ & $42.04$ & $40.39$ & $12.66$ \\
$5166.282$ & $4.13$ & $4.97$ & $0.000$ & $-4.123$ & $10.24$ & $10.36$ & $10.11$ & $1.76$ \\
\hline
\label{tab:feilines}
\end{tabular}
\end{center}
\end{table*}

\subsubsection{Formation}
\label{sec:feformation}

\begin{figure*}[!t]
\begin{center}
	\resizebox{0.49\hsize}{!}{\includegraphics{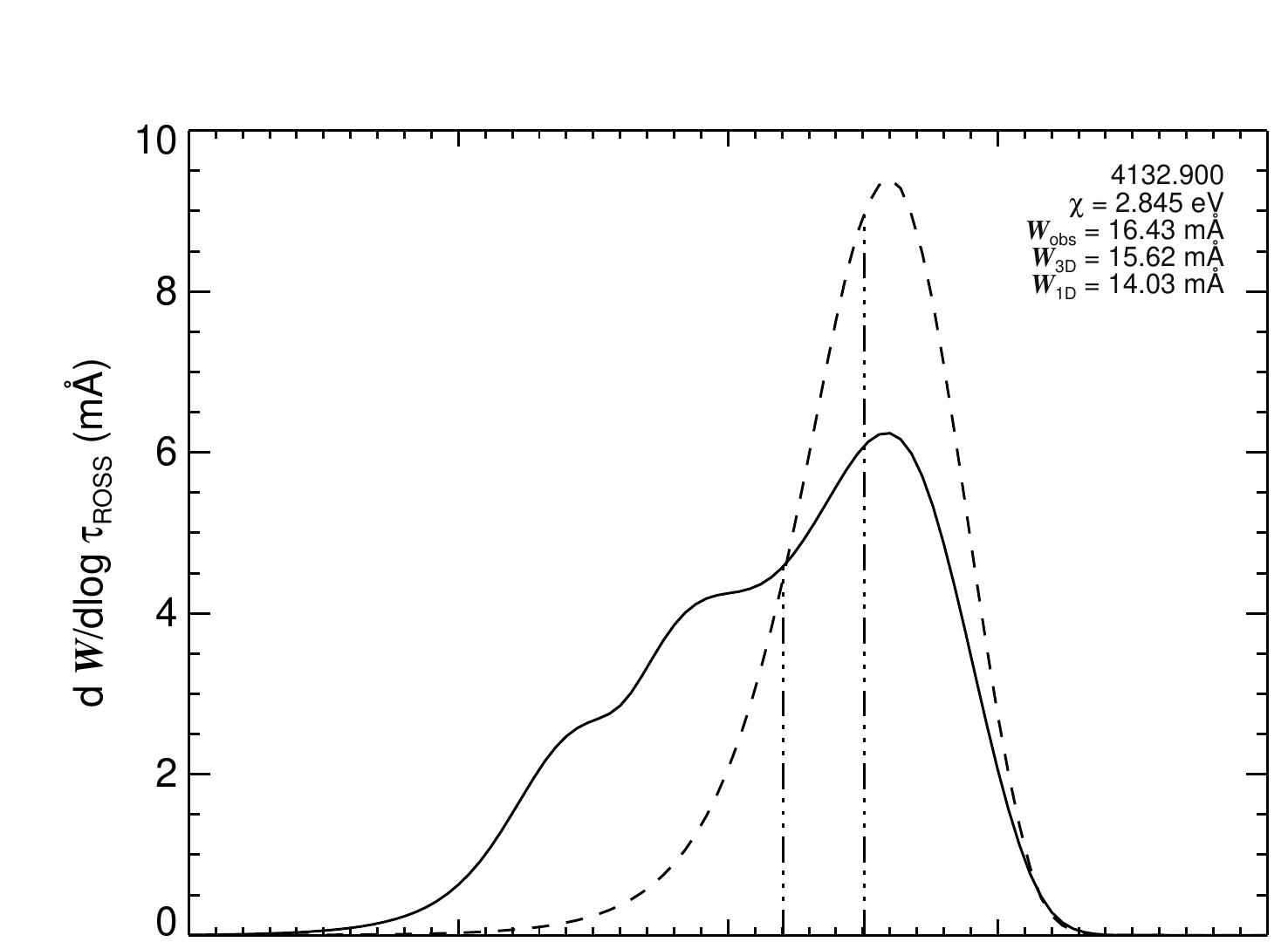}}
	\resizebox{0.49\hsize}{!}{\includegraphics{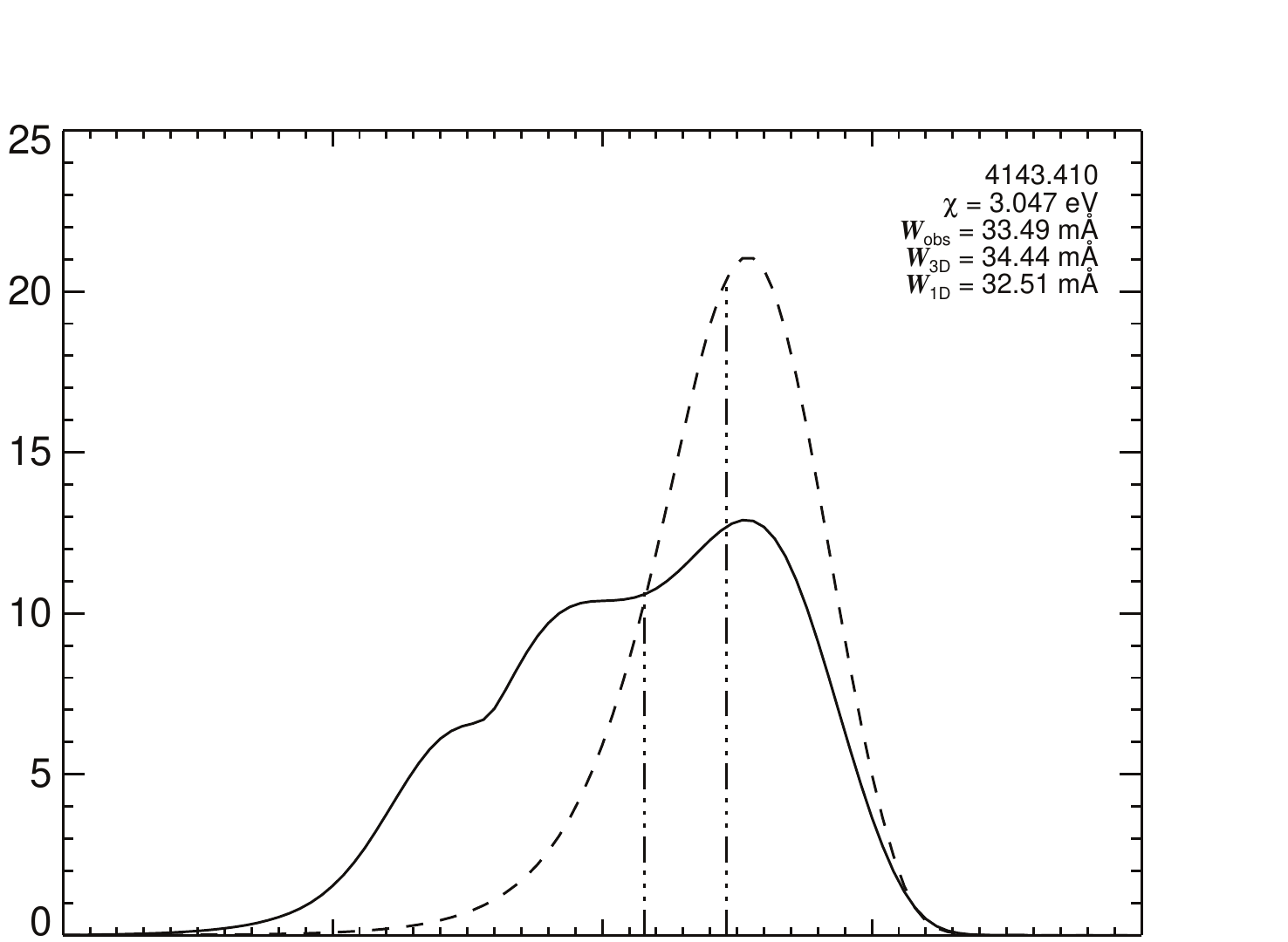}}
	\resizebox{0.49\hsize}{!}{\includegraphics{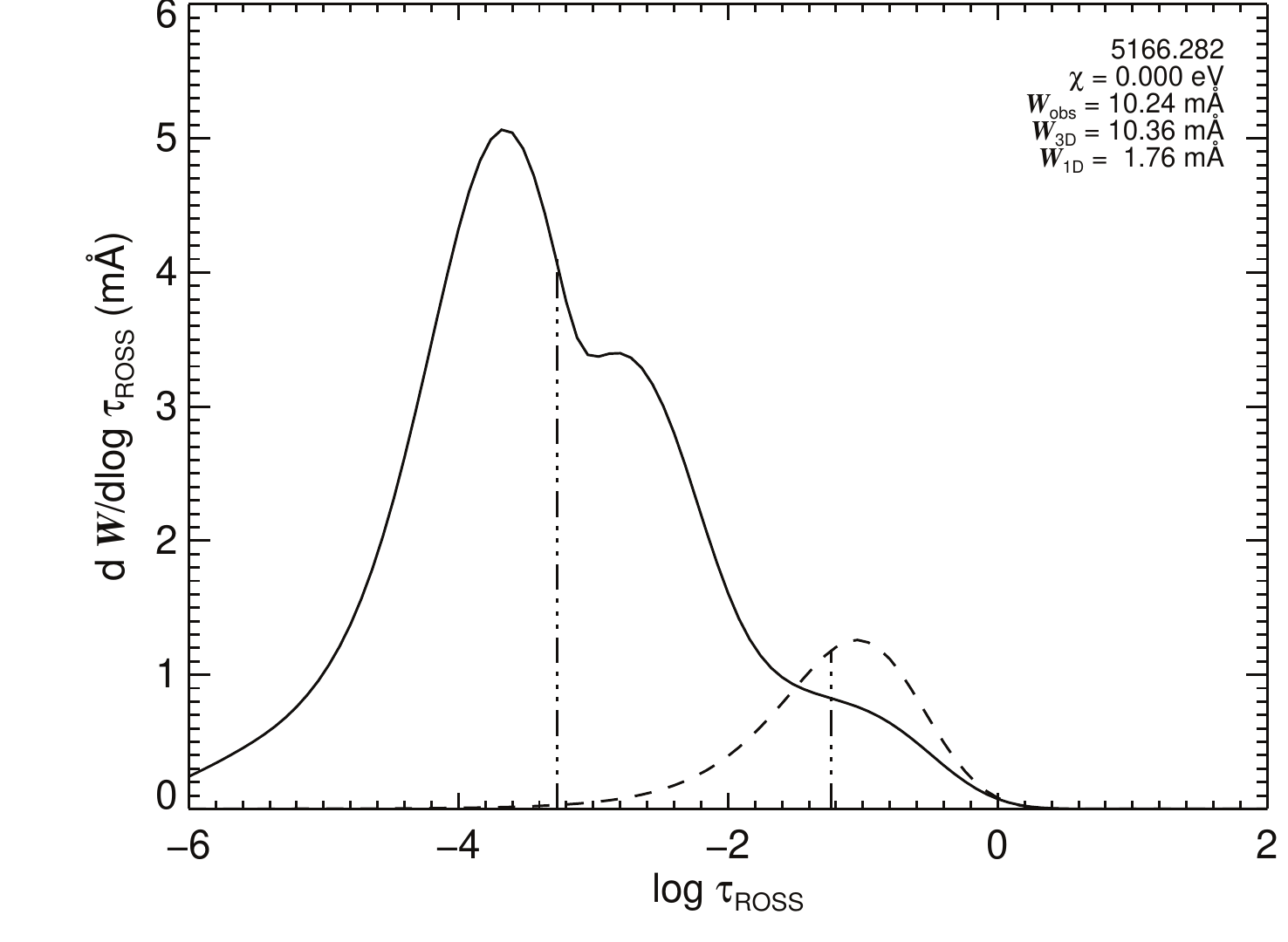}}
	\resizebox{0.49\hsize}{!}{\includegraphics{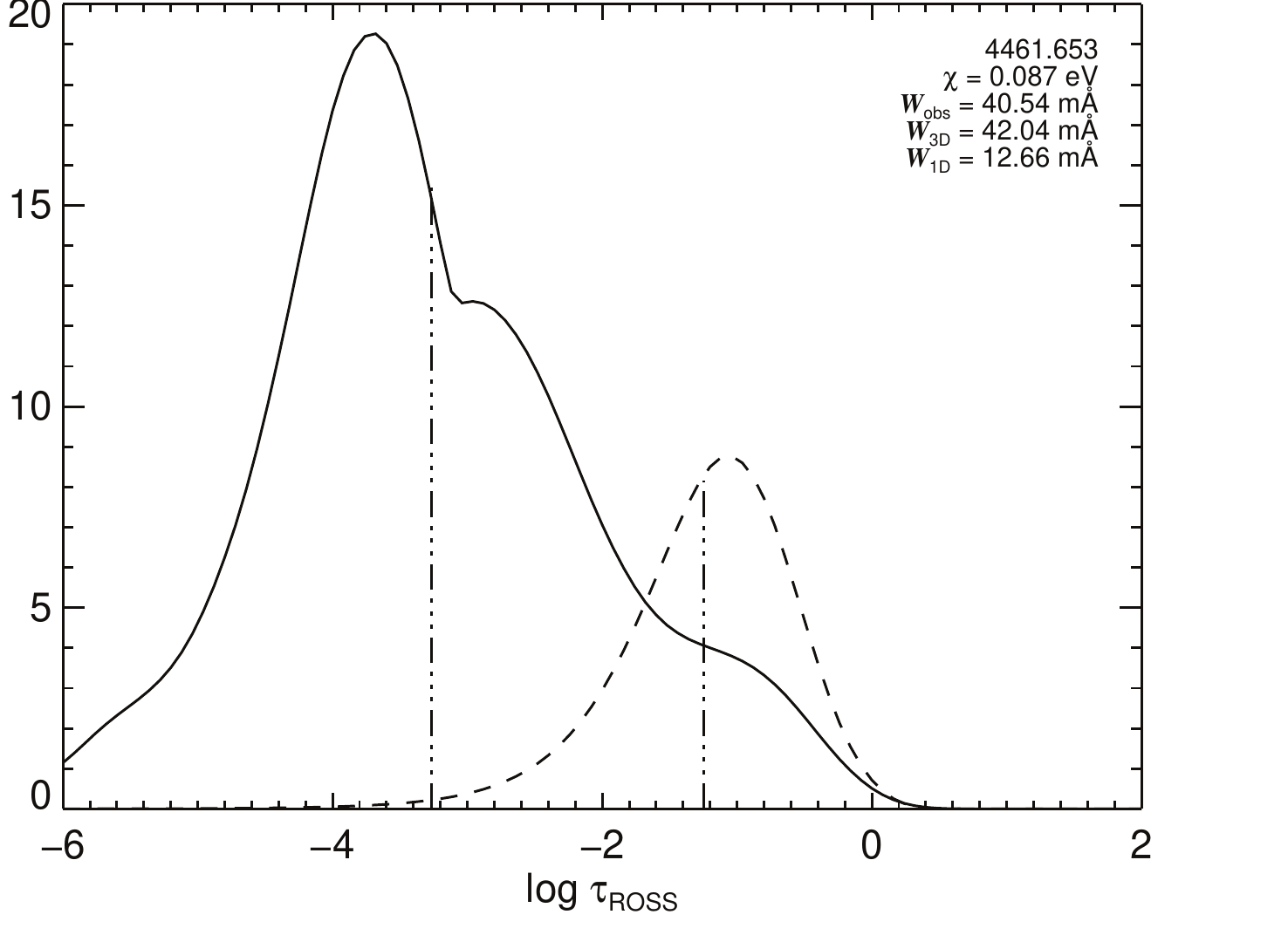}}
	\caption{Equivalent width contribution functions for four distinct iron lines for the 3D (solid line) and counterpart 1D (dashed line) synthesis. Integrals of the curves yield the equivalent width of the \fei\ line. (top left): A high excitation, weak \fei\ line. (top right): A high excitation, strong \fei\ line. (bottom left): A low excitation, weak \fei\ line. (bottom right): A low excitation, strong \fei\ line. These lines were chosen from our sample to illustrate the effect that excitation potential has on formation depths of \fei\ lines. Both the 3D and 1D synthesis were given the same input transition probability ($\loggf$) and excitation potential in each case. While the high excitation lines show good agreement in line strength, the low excitation lines do not. This helps to explain the behaviour shown in Fig.~\ref{fig:chidependence}.}
	\label{fig:fecontfunc}
\end{center}
\end{figure*}

In this section we discuss details of neutral iron line formation for four specific \fei\ lines. They were chosen to represent four extremes of formation that we see in the 81 \fei\ line sample: a high excitation, weak line; a high excitation, stronger line; a low excitation, weak line; and a low excitation, stronger line. The lines selected are listed in Table~\ref{tab:feilines}.

The 3D abundances tabulated in column (2) of Table~\ref{tab:feilines} represent values that best fit the observed profiles, calculated in this investigation. Their resultant equivalent widths are tabulated in column (7). The 1D abundances in column (3) represent values that best fit the observed lines found in the $\chi^2$ analysis conducted in \po, such that their equivalent widths are roughly equal to $W_{\rm obs}$ in the table. Their resultant equivalent widths are tabulated in column (8) under the heading $W_{\rm 1D,1D}$. The equivalent widths tabulated in column (9) under the heading $W_{\rm 1D,3D}$ do not reproduce the 1D abundances listed in column (3) of the table, but instead represent the resultant 1D equivalent widths from the \od\ synthesis for the 3D abundances tabulated in column (2). We note that relative differences between the synthesis produced using \od\ and \kurucz\ atmospheres are immaterial. Therefore, it is of little importance which 1D codes produce the values tabulated, as both are virtually identical. From columns (2) and (3) in Table~\ref{tab:feilines}, it is clear that large discrepancies appear between the 3D and 1D abundances for the two low excitation lines. 

The behaviour of the \fei\ lines can be further explored by examining their contribution functions. In Fig.~\ref{fig:fecontfunc} we have plotted the equivalent width contribution functions over a $\ltaur$ scale for the lines listed in Table~\ref{tab:feilines}. Plotting contribution functions in this way allows for direct comparisons of lines with varying line strength as the $\ltaur$--integral of the contribution functions represent the lines' equivalent widths. For full and precise documentation on their calculation, see \citep{Steffen2010}.

The top left and top right plots show the two high excitation \fei\ lines, $4132.900\,\AA$ (left) and $4143.410\,\AA$ (right). The bottom left and right plots show the low excitation $5166.282\,\AA$ and $4461.653\,\AA$ lines, respectively. The solid lines represent the 3D contribution functions, while the dashed lines represent the 1D contribution functions. The triple dashed-dot lines indicate the depth at which half the equivalent width has been achieved, which we refer to as the characteristic depth of formation. This is a useful value to include as it helps to understand where, on average, the lines are forming in the star. Both the 1D and 3D lines in Fig.~\ref{fig:fecontfunc} have been computed for the same abundances so that their equivalent widths are equal to those in listed in columns (7) and (9) in Table~\ref{tab:feilines}.

Fig.~\ref{fig:fecontfunc} and Table~\ref{tab:feilines} demonstrate that the high excitation lines show little discrepancy in line strength and abundance in the 1D and 3D synthesis, or in characteristic depth of formation, relative to the low excitation lines. The principal difference, as foreshadowed in \S\ref{sec:ionisation}, is that because the 3D models are cooler at lower optical depths (compared to the almost isothermal 1D models), the neutral fraction of iron is several times higher than in the 1D case (Fig.~\ref{fig:ionisation}) and so the \fei\ lines begin to form further out. Consequently the 3D contribution functions extend over a larger range of $\ltaur$, towards smaller values, and this in turn shifts the characteristic formation depth outwards and drive the line strength up. To correct for this, one must decrease the number of absorbers for the low excitation lines to achieve the correct equivalent width, leading to a much lower indicated iron abundance, relative to when they are constructed under 1D.

While the low excitation lines require very little energy to excite the lower levels, which means that even collisions in the cool outermost layers may be sufficient, the high excitation lines require higher energy collisions, which are found only in the deeper layers of the star. The 1D atmosphere cools only from $\sim4800\,{\rm K}$ to $\sim4500\,{\rm K}$ over the interval $-4.0\leq\ltaur\leq-1.5$. Over this interval, the Boltzmann factor $\left(\exp{-\frac{\chi}{kT}}\right)$, which gives the population of for example, a $3\,{\rm eV}$ level relative to a $0\,{\rm eV}$ level, reduces only to $62\%$ of its deeper value. Over this depth range, the change in the relative populations of $0\,{\rm eV}$ and $\sim3\,{\rm eV}$ levels is slight in the 1D model. In contrast, the \tda\ model drops to $\sim3800\,{\rm K}$ at $\ltaur=-4.0$, and the relative population of a $3\,{\rm eV}$ level compared to a $0\,{\rm eV}$ level drops to just $15\%$ of its deeper value. In other words, the ratios of the populations of the higher excitation levels, relative to those of low excitation levels, climbs steeply as you enter the 3D star. They increase by a factor $>6$ in the 3D case, but only by a factor of $1.5$ in the roughly isothermal 1D case. Consequently, the higher neutral iron fraction seen in the outer layers of the 3D star (Fig.~\ref{fig:ionisation}) results in an increase in the opacity in the low excitation lines in the outer layers of the \tda\ star, but crucially not in the high excitation lines since the high excitation levels are not significantly populated until much deeper regions. Thus the high excitation 3D case resembles the 1D case, whereas the low excitation 3D lines benefit from the higher neutral fraction in the cooler outer regions of the 3D models.

While we understand that the different behaviour of the low and high excitation lines in the 1D and 3D models results from the temperature profiles, the fact that we derive such a strong dependence of $\afe$ on excitation potential remains unpalatable, and surely points to shortcomings in the 3D LTE line synthesis presented here. An NLTE calculation for iron lines would almost certainly decrease the neutral iron fraction and alter level populations in the outermost regions, where the low excitation lines are forming according to the current 3D prescription \citep[see][]{Bergemann2012}. Without attempting an NLTE analysis in this work, we do briefly examine the LTE ionisation balance before going on to look at line asymmetries as modelled by the current 3D LTE code. 

The $\vsini$ values determined for all 91 iron lines were compared with the iron line's characteristic depth of formation and no trends were found. Therefore $\vsini$ remains unaffected by this particular 3D effect.

\subsubsection{\ion{Fe}{I} vs. \ion{Fe}{II}}
\label{sec:feivsfeii}
Fig.~\ref{fig:chidependence} shows that substantial abundance differences exist between the $2.5-3.5\,{\rm eV}$ \fei\ and \feii\ lines ($\Delta\afe\sim0.3$\,dex), which could indicate an incorrect atmosphere parameter assignment. Through analysis using the other atmospheres listed in Table~\ref{tab:atmtemp}, it was found that the ionisation equilibrium could be shifted. We have tabulated the abundances found from analysis of the entire sample of \fei\ and \feii\ lines in Table~\ref{tab:feifeiiabundances}, for each atmosphere.

\begin{table}[!t]
\begin{center}
\caption{Iron abundances of the resultant interpolated (${\rm [Fe/H]=-2.5}$) synthesis from every model atmosphere. The corresponding atmosphere details can be found in Table~\ref{tab:atmtemp}. The first row tabulates the results from the high resolution model atmospheres used throughout this section, therefore we have separated it from the rest of the atmosphere sample. Errors tabulated are the standard error in the scatter of the sample.}
\begin{tabular}{l c c c r}
\hline\hline
$\Teff$ & $\logg$ & $\afe$ & $A(\fei)$ & $A(\feii)$  \\
\hline
$5750$ & $3.7$ & $4.78\pm0.02$ & $4.74\pm0.02$ & $5.10\pm0.03$ \\
\vspace{-0.9em}\\
\hdashline
\vspace{-0.8em}\\
$5750$ & $3.7$ & $4.80\pm0.02$ & $4.76\pm0.02$ & $5.12\pm0.03$ \\
$5500$ & $3.5$ & $4.60\pm0.03$ & $4.55\pm0.03$ & $5.01\pm0.03$ \\
$5500$ & $4.0$ & $4.58\pm0.03$ & $4.50\pm0.03$ & $5.24\pm0.03$ \\
$5900$ & $3.5$ & $4.91\pm0.02$ & $4.89\pm0.02$ & $4.99\pm0.03$ \\
$5900$ & $4.0$ & $4.83\pm0.03$ & $4.77\pm0.02$ & $5.25\pm0.03$ \\
\hline
\label{tab:feifeiiabundances}
\end{tabular}
\end{center}
\end{table}

The table shows that the atmosphere with $\Teff=5900\,{\rm K}$ and $\logg=3.5$ shows considerable improvement in the ionisation equilibrium. However, from the 1D analysis, atmospheres with $\Teff=5750\,{\rm K}$ and $\logg=3.7$ showed the closest \fei\ to \feii\ ionisation equilibrium. Naively, it might seem that selection of a different atmosphere parameters is warranted under 3D. However, none of the atmospheres analysed here show any improvement in the excitation potential dependence demonstrated in Fig.~\ref{fig:chidependence}, indicating that iron abundances determined under this 3D LTE paradigm are not currently reliable enough, and the ionisation equilibrium cannot be established adequately. We therefore maintain that the model chosen to best represent HD\,140283 is that defined in the 1D analysis, with $\Teff=5750\,{\rm K}$ and $\logg=3.7$.

\begin{figure*}[!t]
\begin{center}
	\resizebox{0.49\hsize}{!}{\includegraphics{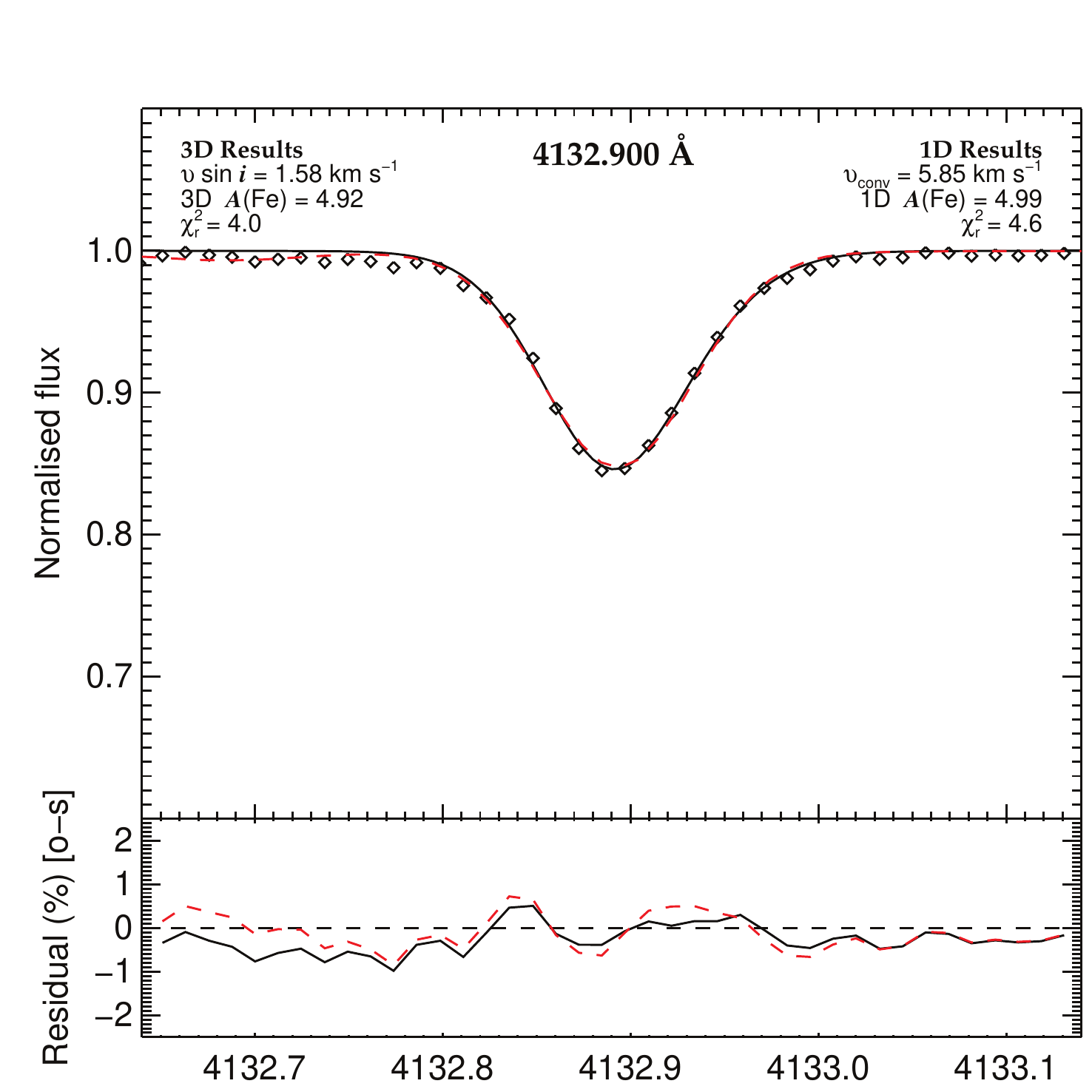}}
	\resizebox{0.49\hsize}{!}{\includegraphics{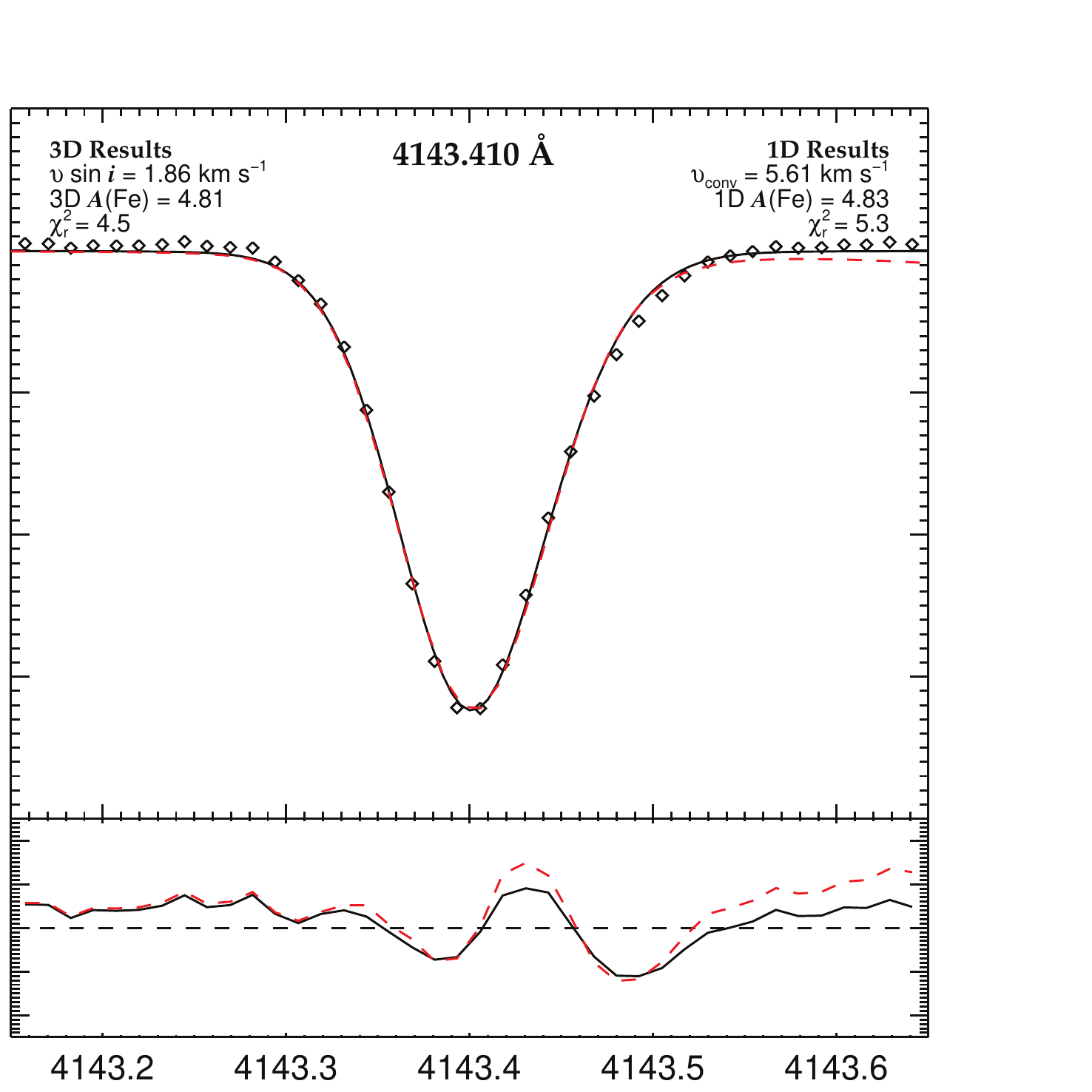}}
	\resizebox{0.49\hsize}{!}{\includegraphics{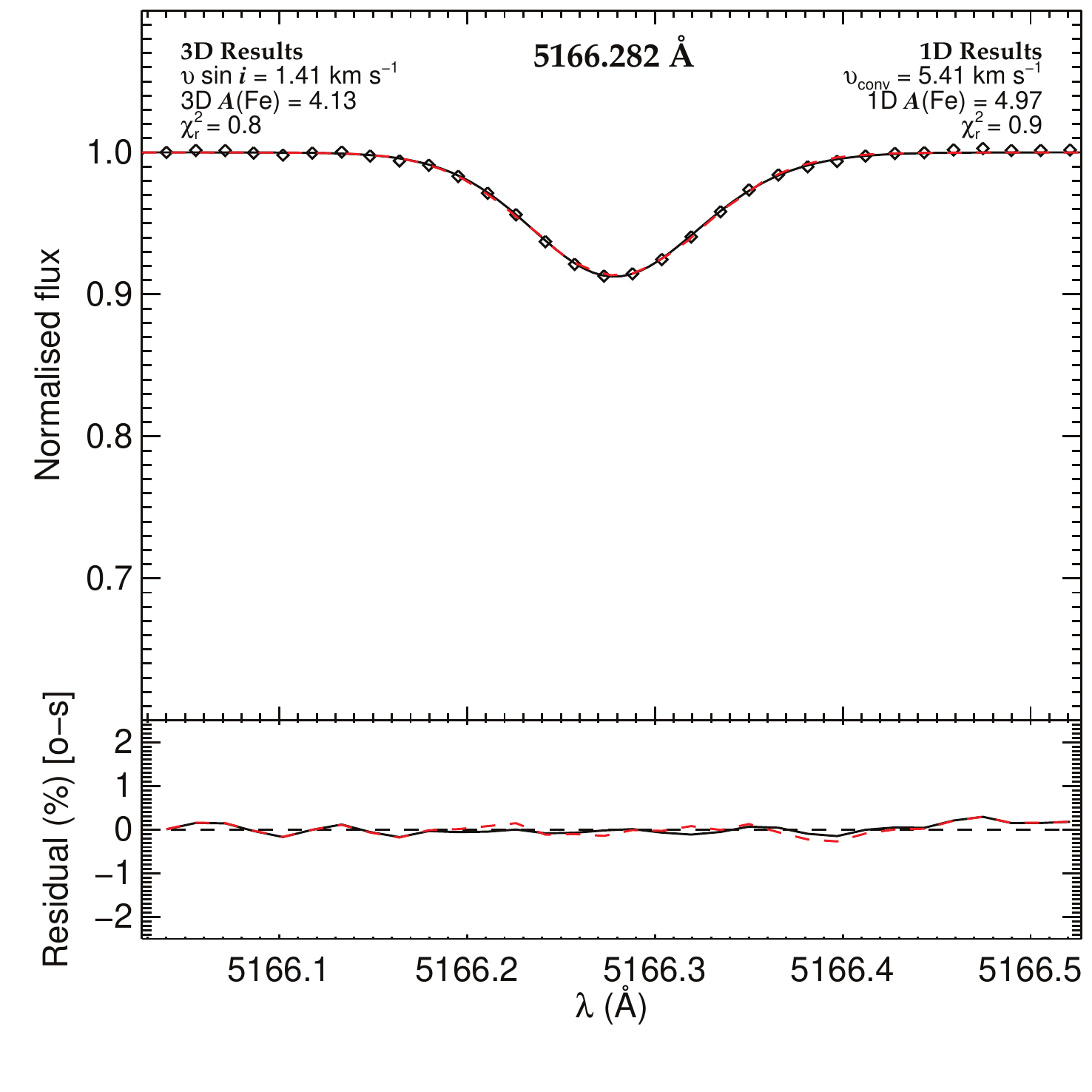}}
	\resizebox{0.49\hsize}{!}{\includegraphics{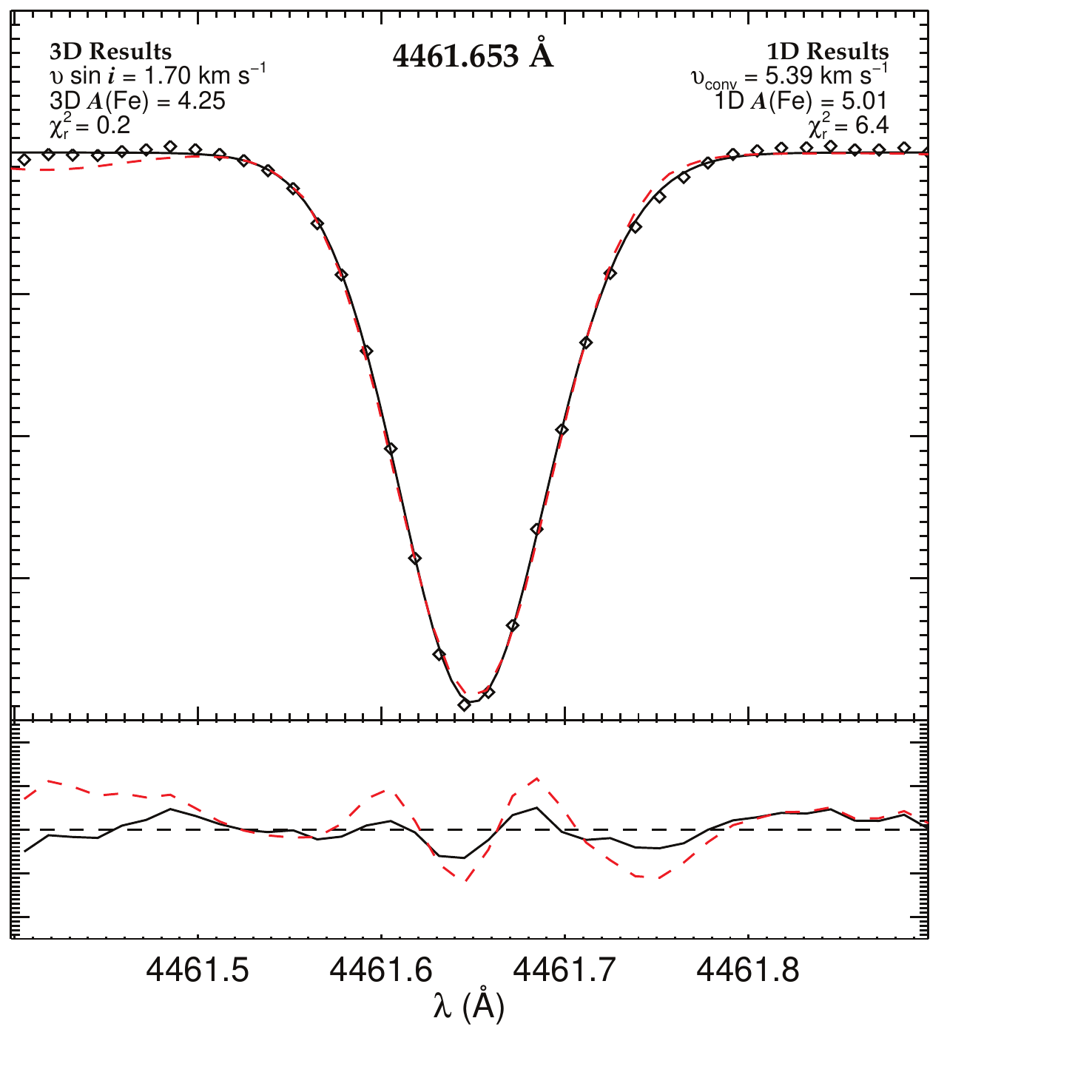}}
	\caption{Best fit 3D (solid black lines) and 1D (red dashed lines) line profiles compared to the observations (diamonds) for the four \fei\ lines listed in Table~\ref{tab:feilines}. A residual plot is shown below each line profile. The same ordering has been used for the plots in this figure as in Fig.~\ref{fig:fecontfunc}, i.e. top panels are the high excitation lines and the bottom panels are the high excitation lines.}
	\label{fig:febf}
\end{center}
\end{figure*}

\subsection{Line asymmetry analysis}
\label{sec:feprofileasymmetry}

One of the main aims of this iron line analysis was to determine whether the 3D synthesis shows notable improvements to the line profile fitting over the 1D synthesis from \po. A large shortcoming of the 1D synthesis from \po\ was the large red-wing asymmetry seen in almost all absorption lines, which we speculated may be due to stellar convection.

Under the 1D paradigm, all line broadening associated with macroturbulent motions in the star (and indeed the star's rotation) are ignored during line synthesis and instead are treated afterwards by convolving a line profile of determined $FWHM$ with the synthesis. In \po\ we used three distinct line profiles: a Gaussian, a radial tangential profile, and a $\vsini$ profile that is used to model stellar rotation. All these methods assume that the broadening, i.e. the redistribution of line opacity caused by these turbulent motions, has a symmetric distribution. The asymmetries seen in the residuals of the synthetic fits to the observed spectrum in \po\ show that this is not the case.

The 3D paradigm is able to approximate the effect of convection in absorption lines as it replicates a dynamic atmosphere, and it is no longer assumed that the redistribution of line opacities is symmetric. 

In Fig.~\ref{fig:febf} we have plotted the four \fei\ lines discussed in \S\ref{sec:feformation}. They are presented in the same order as Fig.~\ref{fig:fecontfunc} to help with direct comparisons. The weak, low excitation line (bottom left panel), is well fit by both paradigms, but there is nevertheless a small improvement found with the 3D fit in the red wing. The high excitation lines (top left and right panels) only show small improvements over the 1D profiles. For the strong, low excitation line (bottom right panel), clear improvements to the line asymmetry in 3D can be seen redward of the line centre, over the equivalent 1D line profile, particularly as judged by the line residuals. In general, similar improvements were found for the entire sample of \fei\ lines.

\begin{figure}[!t]
\begin{center}
	\resizebox{\hsize}{!}{\includegraphics{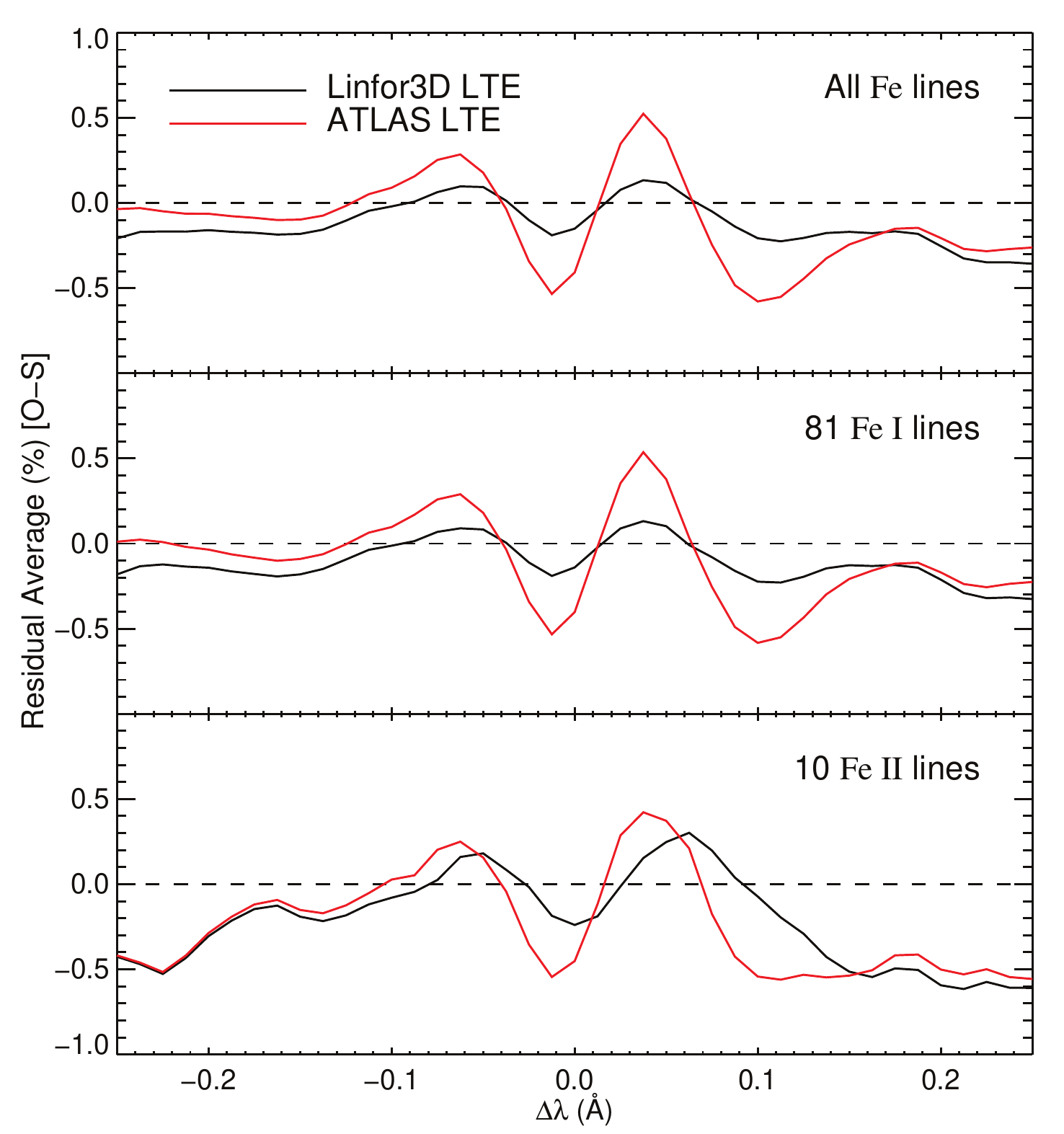}}
	\caption{Reduced noise residual plot for all 91 iron lines (top panel), 81 \fei\ lines (middle panel) and 10 \feii\ lines (bottom panel) for the 1D (red line) and 3D (black line) analyses. Notable improvements can be seen in both the line core and red wing asymmetries under the 3D analysis, which are seen clearly in the 1D analysis. Small differences between the \fei\ and complete iron line sample exist but notable differences between the \feii\ and the \fei\ lines can be seen. However, this is most likely a result of the number of lines in each sample.}
	\label{fig:feresid}
\end{center}
\end{figure}

In Fig.~\ref{fig:feresid} we have plotted averaged, i.e. reduced noise, residuals for the 3D (black line) and 1D (red line) for the entire iron line sample (top panel), \fei\ lines (middle panel) and \feii\ lines (bottom panel) for comparison. The plots were produced by averaging the residuals of all the iron line best fits. This helps remove the random noise contribution found in each individual residual, so that only consistent fitting errors remain. Notable improvements can be seen in the 3D analyses. In particular the asymmetries redward of the line centre that dominate the 1D analysis are severely reduced in the 3D analysis. In general then, it can be said that the 3D paradigm helps to improve the overall fit to an absorption line, as it is better able to reproduce line asymmetries associated with a dynamic stellar atmosphere. This result gives us some reason to believe that the 3D models do provide a more realistic representation of fluid flows in the stellar atmospheres, not withstanding the difficulties discussed in \S\ref{sec:fecorrections}.

\section{Modelling barium and the isotope ratio}
\label{sec:modellingbarium}
In this section we detail the analysis of the singly ionised barium 4554\,\AA\ resonance line. We begin by describing the processes involved in creating the barium line list.

\subsection{Synthesising the barium line}
\label{sec:bariumsynthesis}
Barium has seven principal, stable isotopes. The two lightest isotopes $\left(\element[][130,132]{Ba}\right)$ form via the so-called p-process and have negligible contribution to the overall solar barium abundance. The other five isotopes are formed through a series of neutron-captures via either the s- or r-process mechanisms. The s-process can synthesise all of these isotopes, but shielding by \element[][134,136]{Xe} prevents synthesis of \element[][134,136]{Ba}, by the r-process.

The odd isotopes broaden all barium spectral lines because of hyperfine structure (hfs), in particular the 4554\,\AA\ resonance line. The odd isotopes contribute to the spectral region closer to the wings of the line. The relative strength of the odd isotope lines blueward of the line centre is smaller than those redward of the line centre. When the r-process fraction is large, the odd isotopes contribution to the line profile is increased, and hence the overall line profile becomes shallower, broader and the line asymmetry is increased also. At the other extreme, when barium isotopes form via the s-process, the wing becomes shallower and the core stronger. For weak, unsaturated lines (like the 4554\,\AA\ line in HD\,140283), the isotope configuration has no effect on the overall equivalent width.

Barium lines are synthesised using hfs information from \citet{Wendt1984} and \citet{Villemoes1993}. For detailed information on the line list, we refer the reader to \po.

\subsection{Barium isotope ratio}
\label{sec:bariumisotopes}

\begin{figure}[!t]
\begin{center}
	\resizebox{\hsize}{!}{\includegraphics{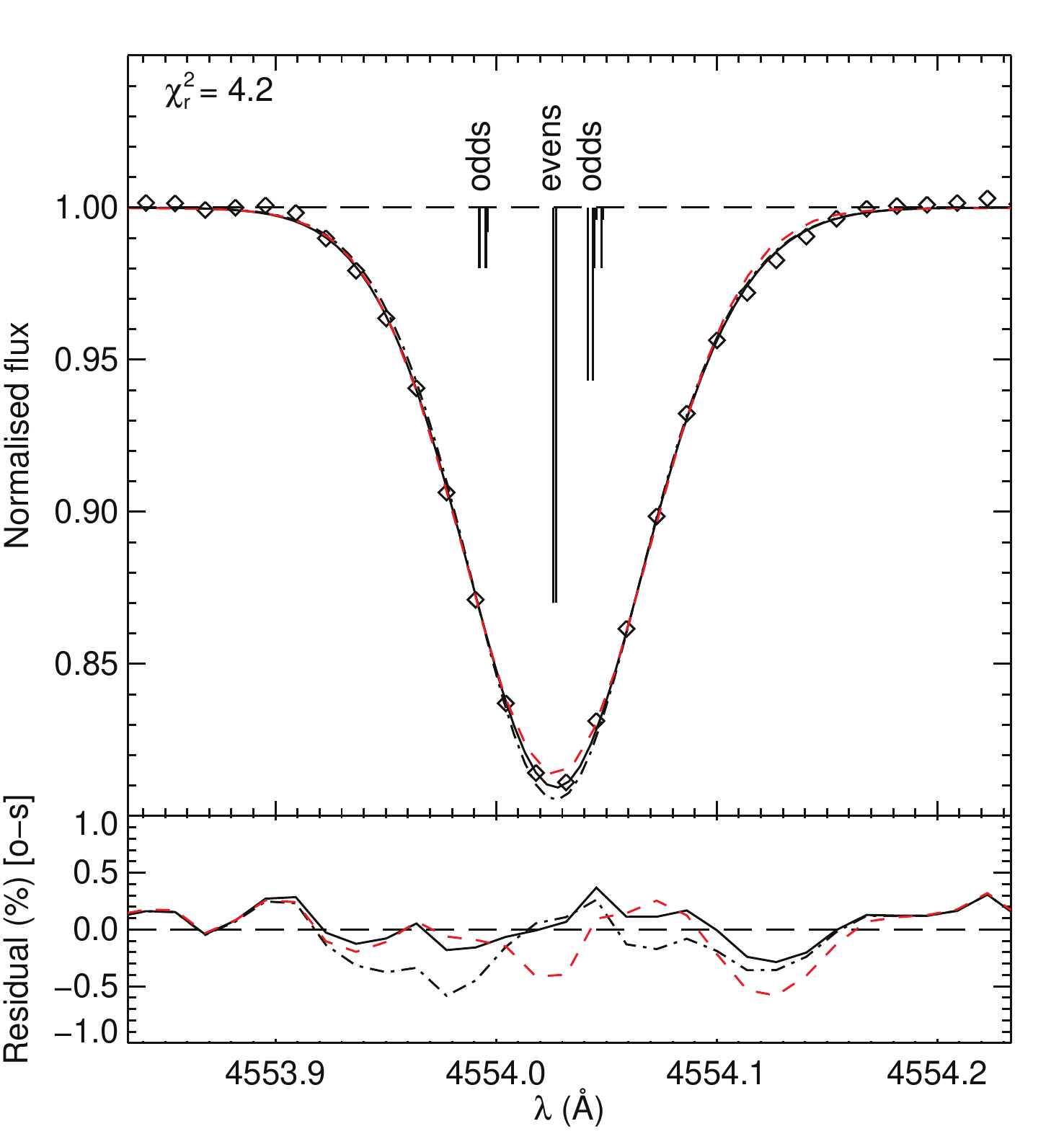}}
	\caption{The best fit 3D (solid black line -- $\fodd=0.38$) and 1D (dashed red line -- $\fodd=0.02$ from {\scriptsize PAPER1}) fits to the observed \baii\ 4554\,\AA\ profile (black diamonds). A residual plot is presented in the bottom panel. We have also included a lower isotope ratio fit to the observed profile (dashed-dot line) for $\fodd=0.25$, i.e. $40\%$, for the same $\vsini$ and $\aba$ values used in the best fit.}
	\label{fig:babestfit}
\end{center}
\end{figure}

The barium isotope ratio, like $\vsini$ from the iron line analysis, was determined using the highest resolution model atmospheres in Table~\ref{tab:atmtemp} labelled as d3t57g37mm20n02 and d3t57g37mm30n02. From the analysis of the iron lines (\S\ref{sec:vsini}), $\vsini=1.65\pm0.05\,\kms$. We convolved this $\vsini$ profile with the barium profile. The instrumental profile, $v_{\rm inst}=3.31\,\kms$, which we model as a Gaussian, was also convolved with the barium profile. Barium profiles were computed with \linfor\ for $11$ barium abundances over a range $-1.54\leq\aba\leq-1.14$ with $\Delta\aba=0.04$. A total of $11$ isotope ratios were synthesised that spanned the entire physical range of isotope configurations, i.e. $\fodd=0.11-0.46$, as set by the nuclear network presented in \citet{Arlandini1999}, with $\Delta\fodd=0.0035$. Synthetic barium profiles were, like the iron profiles, the result of an interpolation of profiles synthesised using the ${\rm [Fe/H]}=-3.0$ and $-2.0$ atmospheres, in each case with abundances set as if ${\rm [Fe/H]}=-2.5$.

\begin{table*}[!t]
\begin{center}
\caption{Values of $\fodd$ calculated from the 20 individual snapshots of the ${\rm [Fe/H]}=-2.0$ and ${\rm [Fe/H]}=-3.0$ atmospheres. The final row of the table lists the averages of each column. The errors reported represent the standard error to the mean.}
\begin{tabular}{l c c c c r}
\hline\hline
Snapshot no. & $\aba$ & $\fodd$  & & $\aba$ & $\fodd$ \\
\cline{2-3} \cline{5-6}
& \multicolumn{2}{c}{${\rm [Fe/H]}=-2.0$}  && \multicolumn{2}{c}{${\rm [Fe/H]}=-3.0$} \\
\hline 
$ 1$ & $-1.38$ & $0.28$ & & $-1.46$ & $0.53$ \\
$ 2$ & $-1.38$ & $0.37$ & & $-1.45$ & $0.40$ \\
$ 3$ & $-1.38$ & $0.35$ & & $-1.47$ & $0.39$ \\
$ 4$ & $-1.38$ & $0.33$ & & $-1.45$ & $0.26$ \\
$ 5$ & $-1.38$ & $0.27$ & & $-1.46$ & $0.35$ \\
$ 6$ & $-1.38$ & $0.46$ & & $-1.47$ & $0.42$ \\
$ 7$ & $-1.38$ & $0.48$ & & $-1.46$ & $0.35$ \\
$ 8$ & $-1.38$ & $0.57$ & & $-1.46$ & $0.47$ \\
$ 9$ & $-1.36$ & $0.37$ & & $-1.47$ & $0.53$ \\
$10$ & $-1.39$ & $0.54$ & & $-1.45$ & $0.20$ \\
$11$ & $-1.36$ & $0.46$ & & $-1.47$ & $0.28$ \\
$12$ & $-1.38$ & $0.39$ & & $-1.47$ & $0.49$ \\
$13$ & $-1.38$ & $0.36$ & & $-1.46$ & $0.59$ \\
$14$ & $-1.37$ & $0.41$ & & $-1.45$ & $0.35$ \\
$15$ & $-1.39$ & $0.32$ & & $-1.45$ & $0.28$ \\
$16$ & $-1.37$ & $0.40$ & & $-1.48$ & $0.57$ \\
$17$ & $-1.38$ & $0.40$ & & $-1.48$ & $0.53$ \\
$18$ & $-1.39$ & $0.37$ & & $-1.46$ & $0.33$ \\
$19$ & $-1.39$ & $0.41$ & & $-1.45$ & $0.19$ \\
$20$ & $-1.38$ & $0.46$ & & $-1.46$ & $0.29$ \\
\vspace{-0.9em}\\
\hdashline
\vspace{-0.8em}\\
\textbf{Average} & $-1.38\pm0.002$ & $0.40\pm0.02$ && $-1.46\pm0.002$ & $0.39\pm0.03$ \\
\hline
\label{tab:snapshot_fodd}
\end{tabular}
\end{center}
\end{table*}

The isotope ratio was determined with a $\chi^2$ routine modified from that of \citet{Perez2009}, which fits synthetic barium profiles with varying barium abundance and isotope configuration. The wavelength shift is also measured by shifting the observed profile so that the best fit can be attained, through the $\chi^2$ minimum. It was found that $\fodd=0.38$, which indicates a contribution by the r-process of $77\%$ \citep[see][Fig.~1 for the $\fodd$ to r-process relationship]{Gallagher2010}.

Fig.~\ref{fig:babestfit} shows the best fit 3D LTE synthetic barium line. Similar to Fig.~\ref{fig:febf}, Fig.~\ref{fig:babestfit} illustrates how the 3D and 1D spectral calculations differ. Not only is the inferred $\fodd$ higher in 3D ($\fodd=0.02\pm0.06$ in \po, as determined from the 4554\,\AA\ and 4934\,\AA\ line analysis), but the residuals to the fit are also improved, both in the core and the red wing, as noted by inspection of Fig.~\ref{fig:babestfit} and the two $\chi^2_{\rm r}$ numbers found in the present work ($\chi^2_{\rm r}=4.2$) and \po\ ($\chi^2_{\rm r}=6.6$).

\subsubsection{Snapshot selection}
\label{sec:snapshots}

In \S\ref{sec:cobold} we described that each atmosphere presented in this paper represents an average of 20 3D structures in time -- snapshots. They were selected from the whole run to represent the statistics of the complete ensemble of snapshots closely. Hence, every profile analysed in the previous sections is produced from 40 individual snapshots: 20 from the ${\rm [Fe/H]}=-2.0$ atmosphere and 20 from the ${\rm [Fe/H]}=-3.0$ atmosphere. In this section we check the influence of the limited statistics and fit individual snapshot profiles to the observed spectrum to determine $\fodd$. The separation of the snapshots in time is long enough that they can be considered as statistically independent samples of the flow field.

Table~\ref{tab:snapshot_fodd} tabulates the values of $\fodd$ and $\aba$ for the 20 individual snapshots for each atmosphere. Synthesis times are unimportant here, but snapshot number 1 and 20 represent the first and last instance in time, respectively. We have used the highest resolution atmospheres that best represent HD\,140283, i.e. $\Teff=5750\,{\rm K}$ and $\logg=3.7$. In this case, $\vsini$ is fixed to the value found from the iron line analysis done with the individual atmospheres ($\vsini=1.89\,\kms$ for the ${\rm [Fe/H]}=-2.0$ atmosphere and $\vsini=1.57\,\kms$ for the ${\rm [Fe/H]}=-3.0$ atmosphere \S\ref{sec:atm_effects}), but is not calculated for individual snapshots because of the desire to see how different snapshots on a single star (fixed $\vsini$) vary. In addition, each snapshot represents an instance in time for the same star, therefore $\vsini$ is no longer a free parameter as it will not vary over time. When the values for $\fodd$ in the table are averaged so that the effective metallicity is ${\rm [Fe/H]}=-2.5$, the resultant value for $\fodd$ agrees with the value reported in \S\ref{sec:bariumisotopes} at the $0.4\sigma$ level. The abundance, when averaged, is identical to that reported in \S\ref{sec:abacorrections}. If $\vsini$ were calculated using the iron lines for each separate snapshot, $\fodd$ would most likely remain constant, given that we see this same behaviour between $\fodd$ and $\vsini$ replicated throughout \S\ref{sec:modellingbarium}.

Table~\ref{tab:snapshot_fodd} indicates an appreciable variation of $\fodd$ from snapshot to snapshot, reflecting temporal variations of the line profile shape. $\fodd$ values encompass almost the entire s- to r-process configuration possible, and in some snapshots $\fodd$ even lies outside the range suggested by current nuclear synthesis models. Therefore it is essential to average over many the snapshots to represent the real star. Table~\ref{tab:snapshot_fodd} shows that the uncertainty of $\fodd$ arising from the small box size is reduced to $\sim0.03$ by averaging over 20 snapshots, which is $\sim10\%$ of the range in $\fodd$ ($0.11$ to $0.46$). The error could be reduced further by computing additional snapshots. However, the given uncertainty is likely an upper limit as the rotational line broadening was not adjusted for each snapshot individually. As we shall see in \S\ref{sec:atm_errors}, a case-by-case adjustment leads to a fairly robust value of $\fodd$.

\subsubsection{Barium abundance differences}
\label{sec:abacorrections}
The barium abundance determined from the best fit synthetic profile is $\aba=-1.43\pm0.05$, where the error assigned represents the propagated uncertainties in the atmospheric parameters (\S\ref{sec:atm_errors}). However, systematic errors encroach on the barium abundance determined, which are due to model selection. It was found that the abundance determined above shifted $\pm0.04$ dex between the ${\rm [Fe/H]}=-2.0$ and $-3.0$ models. Therefore $\aba=-1.43\pm0.05\pm0.04$ as $\vsini$ has no impact on $\aba$. 

In \po\ it was found that $\aba=-1.28\pm0.08$, meaning that an abundance difference of $-0.15$ dex exists between the 3D and 1D best fit profiles. The \baii\ 4554\,\AA\ line is a resonance line. Like the low excitation \fei, lines the 4554\,\AA\ line forms further out in the 3D atmosphere and its contribution function extends over a larger range of $\ltaur$, relative to the 1D atmosphere. It was found that the characteristic depth of formation for the barium line was $\ltaur\sim-2.2$ in 3D, while in 1D $\ltaur\sim-1.5$. It is therefore reasonable to assume that similar abundance effects would affect the barium lines, like we have seen in the iron lines, and that a 3D NLTE analysis of the barium abundance may resolve the issues we outlined in \S\ref{sec:excitation_potential} and \S\ref{sec:feformation}. However, the goal of this work was not to evaluate stellar abundances, but to determine the isotope ratio of barium.

\subsubsection{Random uncertainties}
\label{sec:atm_errors}

In Table~\ref{tab:foddeffects} we have tabulated how $\fodd$ and $\aba$ are affected by changes in temperature and gravity. This was done by synthesising the barium line using each atmosphere listed in Table~\ref{tab:atmtemp} and convolving every barium grid with the corresponding $\vsini$ value calculated through analysis of the iron lines, which were also synthesised using each atmosphere. 

\begin{table}[!t]
\begin{center}
\caption{Atmosphere effects on $\fodd$ and $\aba$. $\vsini$ has been determined for each atmosphere via analysis of the resultant iron line sample synthesis. Results presented here are derived from the interpolated model atmospheres, so that the effective ${\rm [Fe/H]}=-2.5$. The first row presents the results from the high resolution model atmospheres.}
\begin{tabular}{l c c c r}
\hline\hline
$\Teff$ (K) & $\logg$ & $\vsini$ & $\aba$ & $\fodd$ \\
& & ($\kms$) & & \\
\hline
$5750$ & $3.7$ & $1.65\pm0.05$ & $-1.43$ & $0.38$ \\
\vspace{-0.9em}\\
\hdashline
\vspace{-0.8em}\\
$5750$ & $3.7$ & $2.09\pm0.05$ & $-1.41$ & $0.35$ \\
$5500$ & $3.5$ & $2.49\pm0.04$ & $-1.63$ & $0.32$ \\
$5500$ & $4.0$ & $3.19\pm0.03$ & $-1.49$ & $0.33$ \\
$5900$ & $3.5$ & $0.93\pm0.05$ & $-1.43$ & $0.31$ \\
$5900$ & $4.0$ & $2.51\pm0.04$ & $-1.31$ & $0.33$ \\
\hline
\label{tab:foddeffects}
\end{tabular}
\end{center}
\end{table}

It is shown that $\fodd$ is not very sensitive to uncertainties associated with temperature and gravity. In \po\ we state that the uncertainty relating to the temperature and $\logg$ of HD\,140283 were $\sigma_{T}=\pm100\,{\rm K}$ and $\sigma_{\logg}=\pm0.1$. We adopt these uncertainties here as well. The uncertainty associated with the gravity of HD\,140283 means that $\sigma_{\fodd,\logg}=0.003$ and the temperature uncertainty gives $\sigma_{\fodd,T}=0.003$. In summary, the resultant uncertainty to $\fodd$, derived from the sensitivity to the atmosphere parameters ($T,\logg$), is negligible at $\pm0.004$. Measuring how varying $\vsini$ directly affects $\fodd$ allowed us to determine that $\delta\fodd/\delta\vsini=-0.31\,(\kms)^{-1}$. This implies that $\sigma_{\fodd,\vsini}=0.02$, as the scatter in $\vsini$ was found to be $\pm0.05\,\kms$ (\S\ref{sec:vsini_result}). Therefore, the total random error we assign to $\fodd$, derived from $\vsini$, temperature and $\logg$ uncertainties, is $\pm0.02$.

For the barium abundance, we find that $\sigma_{\aba,T}=0.05$ and $\sigma_{\aba,\logg}=0.02$, meaning that we assign $\aba$ the error $\pm0.05$. We note that because of the effects outlined in \S\ref{sec:excitation_potential} and \S\ref{sec:feformation}, this error is most likely not a true representation of the total assigned error on the barium abundance.

It would appear that $\fodd$ is more sensitive to model resolution, as judged by the differences in the high and low resolution model results tabulated in rows (1) and (2), respectively, but this is a systematic effect, which we now move on to discuss.

\subsubsection{Systematic uncertainties}
\label{sec:systematics}
In Table~\ref{tab:foddeffects} it is shown that $\vsini$ varies quite considerably when atmosphere parameters are changed. We find an opposite trend between the atmospheric velocity amplitudes in the 3D models and the derived $\vsini$. This suggests that these changes are the result of $\vsini$ compensating for the changes in the broadening by the fluid motions. While this has no immediate impact on $\fodd$, as it remains fairly consistent in Table~\ref{tab:foddeffects}, it is useful to quantify the effect that the atmospheric parameters have on $\vsini$. By error propagation, it is found that the systematic error contribution by $\vsini$ on $\fodd$ is $\pm0.12$. One could conservatively adopt $\pm0.12$ as an indicative systematic error for an independently derived $\vsini$. However, this would most likely over estimate the total systematic error to $\fodd$, as one would independently derive $\vsini$ values for each atmosphere used to calculate $\fodd$, to adjust for differences in how the velocity fields are treated under different model atmosphere parameters. Thus the systematic error is reduced to the scatter in the results for $\fodd$, tabulated in Table~\ref{tab:foddeffects}. Therefore, we adopt the systematic error based upon this, which is $\pm0.02$.

Finally, Table~\ref{tab:foddeffects} shows that there is a systematic uncertainty associated with model resolution, although more data would be required to make an accurate uncertainty assessment. Nevertheless, we estimate the error on $\fodd$ to be $0.03$, based on what we have presented here. The total systematic error on $\fodd$, derived from the effects of snapshot selection ($0.03$ \& $0.02$ from Table~\ref{tab:snapshot_fodd}), model parameters ($0.02$) and model resolution ($0.03$), is $\pm0.06$.

\subsection{Result}
\label{sec:fodd}
We find that $\fodd=0.38\pm0.02\pm0.06$, which means that barium shows an r-process signature of $77\pm6\pm17\%$ in HD\,140283. Errors presented represent the total assigned random (\S\ref{sec:vsini_result} and \S\ref{sec:atm_errors}) and systematic errors (\S\ref{sec:systematics}), respectively. This not only suggests that barium isotopes in HD\,140283 have a predominantly r-process origin, but also strengthens our speculation in \po\ that the classical 1D LTE prescription is not adequate for the task of determining isotope ratios, using the line asymmetry method presented in this paper and in \po. We are not suggesting that 1D LTE modelling techniques are no longer viable tools for stellar atmosphere analyses, but rather we are highlighting a limitation of the paradigm.

\subsection{Discussion}
\label{sec:discussion}
In \po\ it was speculated that a 3D treatment of the convection would possibly increase the s-process fraction as found by \citet{Collet2009}. Clearly the result found here does not agree with their determination of $\fodd$ and indeed with that speculation made in \po. However, it would seem from their $\vsini$ determination that the atmosphere chosen to model HD\,140283 may not describe the fluid flows of the atmosphere well, which could be the result of several parameters: the numerical resolution of the model is not high enough (we see a similar increase in $\vsini$ from our lower resolution models); the temperature of the model was too low (leading to smaller convective velocities); too few opacity bins were used in the model (changing the overall structure of the atmosphere). One or all of these would result in a larger $\vsini$ value to compensate. 

Additionally, we note that the iron lines examined by \citet{Collet2009} are, in general, stronger on average than those used in this investigation. We have four iron lines in common with \citeauthor{Collet2009} Of these four lines, two were weak lines, and one of these was an \feii\ line. However, the two stronger iron lines in the sample show a higher than average $\vsini$ value, which is a common feature seen in the stronger lines, as shown in Fig.~\ref{fig:vsiniEW}. While two lines are not a basis for strong evidence, this could suggest that the trend $\vsini$ has with line strength (seen in Fig.~\ref{fig:vsiniEW}) is not random. If this is the case, then this would suggest that a bias in $\vsini$ exists (when synthesising under a 3D LTE paradigm), which will depend on the selection criterion used; the stronger the lines synthesised under 3D LTE, the larger the $\vsini$ value.

Finally, \citet{Lind2013a}, who measure lithium isotopes in HD\,140283, report $\vsini=2.83\pm0.03\,\kms$ under LTE, comparable to that of \citet{Collet2009}. However, the \citeauthor{Lind2013a} published $\vsini$ value they assign under the LTE paradigm had been artificially increased to compensate for reduction in the instrumental broadening $FWHM$. The actual $\vsini$ value is $\vsini=2.04\pm0.03\,\kms$ \citep{Lind2013b}. This value agrees at the $0.8\sigma$ level with our $\vsini$ determination and at the $1.8\sigma$ level with that of \citet{Collet2009}.

In order to help understand the Ba isotopic ratios derived using 1D models, we experimented by fitting pure-s and pure-r 1D synthetic spectra not to noisy observational data but to two noise-free 3D synthetic spectra, likewise calculated for pure-s and pure-r isotope mixes. The 3D synthetic spectra were thus treated heuristically as observational spectra of infinite $S/N$, and small offsets in the wavelength, strength and broadening of the synthetic 1D profiles were permitted, as was the case when fitting the 1D profiles to real observations.

Interestingly, for both isotopic mixes the 3D profiles were better fit by the 1D pure-s profiles. In other words, if the 3D synthetic spectra are reasonable approximations to real stellar spectra, then this experiment suggests that the subtle differences between pure-s and pure-r real spectra are much smaller than the differences between 1D and 3D spectra, to the extent that fitting 1D spectra cannot recover the correct isotopic ratio; such fits are dominated by the gross 1D vs. 3D profile differences rather than the subtle s- vs. r-process profile differences.

This experiment does not demonstrate that current 3D synthetic spectra {\it are} a better representation of real spectra, but if they are then it helps us understand why 1D modelling of a star whose heavy-element ratios are r-process-like can lead to 1D-derived isotopic signatures which appear s-process-like. It also underscores the importance of solving remaining issues with 3D modelling, such as the dependence of abundance on excitation potential discussed extensively in \S\ref{sec:excitation_potential}.

\section{Conclusions}
\label{sec:conclusion}
We find that the barium isotopes in HD\,140283 show an r-process origin as $\fodd=0.38\pm0.02\pm0.06$ ($77\pm6\pm17\%$ r-process), where the errors represent random and systematic errors, respectively. This result contradicts 1D the result presented in \po, where the isotope configuration was shown to have a distinctly s-process origin, such that $\fodd=0.02\pm0.06$. It is important to note that the 1D $\fodd$ value reported from \po\ is the result of inverse-variance-weighting the isotope ratios from the 4554\,\AA\ ($0.01\pm0.06$) and 4934\,\AA\ ($0.11\pm0.19$) lines, which were both measured. The differences between the two results are most likely due to the way in which a 3D and 1D model atmosphere simulate fluid flows in a star. In a 1D model, fluid flow is approximated by the fudge factor, macroturbulence, which must be independently measured so that $\fodd$ can be determined. It was established in \po\ that determining macroturbulence can lead to problems when measuring $\fodd$ as we reported that $\delta\fodd/\delta v_{FWHM}=-0.7\,(\kms)^{-1}$. For a 3D atmosphere, macroturbulence is redundant though $\vsini$ must be derived instead. As a 3D atmosphere is dynamic, it is better able to the replicate velocity fields of convective cells. Importantly, these fluid flows are not symmetric, which leads to asymmetrically broadened line profiles.

The isotopic composition of barium behaves like an asymmetric broadening mechanism -- the larger the r-process signature, the more asymmetric the line becomes. Therefore $\fodd$ is extremely sensitive to external broadening mechanisms, in particular the macroturbulence parameter in a 1D analysis and $\vsini$ in 3D. Because macroturbulence is measured using a different species of line, often iron lines, the macroturbulence will also be under or over estimated, impacting the isotope ratio, and hence the value for $\fodd$. In addtion, because lines are broadened symmetrically with a 1D prescription, asymmetric convective effects are inevitably mistaken as an isotope signature, leading to incorrect estimates of the r-process contribution of $\fodd$.

It was also demonstrated in \S\ref{sec:discussion} that 3D barium line profiles, for either an s-process-only or r-process-only isotope mixture, are better fit by 1D s-process-only synthetic profiles. This may help to explain why $\fodd$ was determined to have an s-process-only isotope mixture in \po, if 3D line profiles better represent real stellar spectra.
 
Our result also contradicts the result published by \citet{Collet2009}, who also employ a 3D modelling technique to calculate $\fodd$ in HD\,140283. However, we speculate in \S\ref{sec:discussion} that this apparent difference in isotope configuration could be caused by a number of differences between the model atmospheres used by them and in the present analysis, or by a deeper routed problem, caused by inherent inadequacies in the 3D LTE line synthesis, though we stress here that this is purely speculative at this moment. Nevertheless, it is evident that $\vsini$, like the macroturbulence in a 1D investigation, is extremely important when measuring $\fodd$.

We have also shown that in LTE, our 3D model does not provide a unique abundance of iron as derived from \fei\ lines. The abundances as derived from different lines have a large dependence on their excitation potential, caused ultimately by the 3D atmosphere models' temperature structure. We strongly suspect that an NLTE line formation treatment will resolve this issue. Currently, robust NLTE calculations for iron are at the limit of what is computationally feasible. However, \citet{Shchukina2005a} and \citet{Bergemann2012} made attempts to tackle the 3D NLTE problem for iron in HD\,140283 by examining line profiles produced with 1.5D NLTE and \tda\ NLTE line synthesis, respectively. Like us, they find a strong trend of the abundances from \fei\ lines in excitation potential in LTE; in NLTE the trend vanishes -- in fact is even overcompensated for -- making it very plausible that NLTE effects are the culprit of the present shortcomings in the abundance and $\vsini$ results we report.

In comparison to 1D, the fits to the observed iron and barium profiles were, in general, improved under the 3D paradigm. Moreover, we demonstrated that changes to the model atmosphere parameters and resolution have little effect on $\fodd$, making our result quite robust. However, further work is in order to improve the line formation calculations in the 3D models, and to clarify the reasons for the conflicting results on the barium isotopic composition in HD\,140283 reported in the literature.

\begin{acknowledgements}
AJG acknowledges the members of the CIFIST collaboration for access to the CIFIST \cobold\ atmosphere grids. HGL acknowledges financial support by the Sonderforschungsbereich SFB 881 ``The Milky Way System'' (subproject A4) of the German Research Foundation (DFG).
\end{acknowledgements}

\nocite{*}
\bibliographystyle{aa}
\bibliography{20563}
\end{document}